\documentclass[final,5p,times]{elsarticle}

\usepackage{graphicx}
\usepackage{amsmath,amssymb,amsfonts}
\usepackage{amsthm}
\usepackage{tabularx}
\usepackage{subcaption}
\usepackage{booktabs}
\usepackage{float}

\usepackage{array}
\usepackage{ragged2e}
\newcolumntype{Y}{>{\RaggedRight\arraybackslash}X}
\usepackage{lscape}
\usepackage{adjustbox}
\usepackage{enumitem}
\usepackage{xurl}
\usepackage[hidelinks]{hyperref}


\begin{document}

\begin{frontmatter}

\title{An Empirical Security Analysis of Open-Source Software Used in Onboard Satellite Systems}

\author[aff1]{Roee Idan\corref{cor1}}
\ead{roeeidan@post.bgu.ac.il}
\author[aff1]{Tomer Cohen Galor}
\author[aff1]{Asaf Shabtai}
\author[aff1]{Yuval Elovici}

\cortext[cor1]{Corresponding author}

\address[aff1]{Faculty of Computer and Information Science, Ben-Gurion University of the Negev, Beer-Sheva 8410501, Israel}

\begin{abstract}
The use of open-source software (OSS) in satellite flight systems is increasing as missions adopt reusable frameworks, shared libraries, and community-maintained components.
While this accelerates development, it also introduces software-security risks into systems where patching is costly and failures may affect mission operations.
This paper presents an empirical security study of OSS used in onboard satellite systems.
We analyze 126 public repositories using a pipeline that combines software bill of materials generation, software composition analysis, static application security testing, infrastructure-as-code analysis, and secret scanning.
After rule-based cleaning, onboard-scope filtering, and fingerprint-based deduplication, the pipeline produced a final dataset of 2,827 findings.

The results show that security findings are widespread but unevenly distributed.
Medium-severity findings account for 49\% of the dataset, and 72\% are classified as medium severity or higher.
A Common Weakness Enumeration (CWE)-based taxonomy assigns all findings to eight weakness families.
Memory Safety and Code Quality dominate the dataset, followed by Input Validation and Injection.
Most findings occur in project-developed code, accounting for 81.4\% of the dataset, while external dependency code remains a relevant source of findings.
While these findings do not establish mission-specific exploitability, they provide an empirical characterization of recurring security patterns across the open-source onboard satellite software ecosystem, helping quantify their prevalence and prioritize areas that warrant the greatest security attention.
\end{abstract}

\begin{keyword}
Open-source software, Satellite cybersecurity, Onboard software, Flight software, Static analysis, Software composition analysis, Supply-chain security
\end{keyword}

\end{frontmatter}

\section{Introduction}
\label{sec:intro}

Over the past decade, the number of satellites launched and operated has increased rapidly, growing from dozens to thousands~\cite{yue2023low,cratere2024board}.
This growth is most visible in low Earth orbit (LEO) constellations and CubeSat-class platforms, and has been enabled by declining launch costs and expanded rideshare opportunities~\cite{yue2023low}.
Satellite systems are used in academic, commercial, government, and military settings for a wide range of purposes, including communications, Earth observation, navigation, and scientific exploration~\cite{cratere2024board}.
As satellite development and operation expand across private industry, public agencies, universities, and research institutions, open-source software (OSS) is increasingly incorporated into mission software stacks.

OSS offers practical advantages for satellite development, including faster prototyping, reduced cost, reuse of existing components, and access to publicly inspectable code across mission software stacks~\cite{curbo2023research}.
This practice is also supported by major space agencies.
NASA publishes and reuses open-source components, most prominently the core Flight System (cFS), which has been used on multiple missions~\cite{nasa2025cfs}, while the European Space Agency (ESA) curates open-source resources that support space software development and operational applications~\cite{esa_oss_resources}.
In parallel, standardized form factors and commercial off-the-shelf components further compress development cycles and lower barriers to entry.
As a result, OSS is increasingly present in space software pipelines, and its adoption continues to grow as mission operators and space agencies standardize reusable frameworks, shared libraries, and development tooling; in some cases, entire flight-software repositories are publicly available~\cite{curbo2023research}.

However, the use of OSS also expands the satellite software supply chain and introduces cybersecurity risks that are difficult to manage in onboard systems.
Indirect and transitive dependencies can import vulnerabilities that originate far outside the space domain~\cite{ohm2020backstabber,ladisa2024understanding}.
Package ecosystems, build scripts, and update channels can therefore become attack surfaces, creating exposure that may persist across development, integration, and mission operation~\cite{ohm2020backstabber,kampourakis2025cracks,andreoli2023prevalence}.
Beyond inherited dependency flaws, OSS repositories may also contain local weaknesses such as security misconfigurations, hardcoded secrets, unsafe coding patterns, and unmaintained components with known Common Vulnerabilities and Exposures (CVEs).

In spacecraft, OSS risks are amplified by operational constraints.
Software updates must contend with intermittent communication links, strict safety assurance requirements, and lengthy validation cycles, so weaknesses present at launch may remain unpatched for years~\cite{khan2024space,bace2024securing}.
Mission operators may also defer changes when the operational risk of an update exceeds the perceived security benefit.
These constraints make vulnerability management more difficult than in conventional software environments, where patches can often be tested and deployed more rapidly.
Together, long mission lifetimes and limited patch windows can transform common software weaknesses into mission risk~\cite{lee2023vulnerabilities}.

Researchers have begun to document cybersecurity weaknesses in satellite systems.
Empirical analyses have identified exploitable conditions in deployed satellite firmware and ground software~\cite{willbold2023space,jero2024securing}, while studies of university-built satellites report recurring issues such as insecure dependencies and weak authentication~\cite{mcamis2025short}.
Other work has examined emerging attack surfaces in autonomous and software-defined satellite systems, as well as the broader security implications of rapidly expanding satellite deployments~\cite{weber2024space,yue2023low}.
However, prior work on onboard satellite security has mainly examined individual systems, specific subsystems, or conceptual attack surfaces~\cite{curbo2023research,curbo2024attack}.
To the best of our knowledge, no prior study has systematically examined how security weaknesses are distributed across open-source software repositories used in onboard satellite systems.

This paper presents an empirical study of the security posture of OSS used in onboard satellite systems.
We curated and analyzed 126 public repositories containing onboard-relevant satellite software, with the goal of characterizing ecosystem-level security findings rather than evaluating or ranking individual projects.
We examine how security findings are distributed across repositories, severity levels, onboard subsystems, code sources, and weakness categories, assess onboard software update capability, and map the findings to SPARTA's Threats to Space Systems taxonomy.
The goal is to raise awareness and provide concrete suggestions for improving the security of OSS for onboard satellites, thereby reducing mission risk and making OSS safer for wider use.

Specifically, we provide a high-level overview of security findings in OSS used in onboard satellite systems, focusing on their prevalence, severity, repository distribution, subsystem distribution, code source, and weakness categories.
To that end, we designed and applied an analysis pipeline for this study.
The pipeline includes software bill of materials (SBOM) generation to enumerate dependencies, software composition analysis (SCA) to identify known vulnerable components, static application security testing (SAST) to detect code-level weaknesses, infrastructure-as-code (IaC) checks, and secret scanning.
We then merged and deduplicated the outputs into a consolidated dataset.
Together, these analyses yield a unified view of security-relevant findings across OSS used in onboard satellite systems.

We find that security findings are widespread but unevenly distributed across the studied repositories.
Of the 126 repositories analyzed, 92 contain at least one retained finding.
The analysis produced a final dataset of 2,827 findings, of which 72\% are classified as medium severity or higher.
Memory Safety and Code Quality are the dominant weakness families, followed by Input Validation and Injection.
Nearly half of the findings occur in OBC, CDH, and flight software (48.2\%).
Most findings occur in project-developed code (81.4\%), showing that the observed findings are concentrated in local implementation weaknesses while inherited dependency risks remain relevant.
We conclude with practical mitigation recommendations to support safer use of OSS before flight and throughout mission operations.
\section{Background and Related Work} \label{sec:background}

\subsection{Space Mission Software and Constraints}

Modern space missions are software defined, relying on extensive software stacks across both ground and flight segments~\cite{thibault2022spaceflight}.
Onboard software operates in a constrained and safety critical environment.
This stems from the high cost of launch per kilogram, which limits spacecraft mass, and from the need for all components to withstand harsh space conditions, including radiation and extreme temperature fluctuations.
Together, these physical and financial pressures result in platforms with limited computational capacity, memory, and power~\cite{cratere2024board,yue2023low}.

The consequences of software failure are also more severe in space.
Faults in onboard software can degrade mission functionality, interrupt communications, damage payload operations, or in extreme cases contribute to the loss of a spacecraft.
Core onboard software includes flight software (FSW), which supports command and data handling (C\&DH), hardware abstraction, telemetry, telecommand, and mission control functions, as well as payload application software that manages instruments and mission specific operations~\cite{cratere2024board}.

Operational realities in space complicate maintenance and patching, making post deployment remediation fundamentally different from terrestrial operations~\cite{bace2024securing}.
For deployed satellites, software updates are high risk events, constrained by intermittent communication links, limited uplink bandwidth, and strict scheduling that prioritizes mission safety and availability~\cite{lee2023vulnerabilities}.
Any change requires extensive ground based verification before being uplinked, and operators must weigh the operational risk of applying an update against the security risk of leaving a vulnerability unpatched.
Because missions range from a few years to multiple decades~\cite{khan2024space}, vulnerabilities present at launch can persist for years, turning otherwise routine software defects into long term mission liabilities~\cite{sharmin2025cyber}.

\subsection{Open-Source Software in the Space Sector}

OSS has become a strategic enabler in the modern space sector.
This shift is driven by the need to lower costs, accelerate development, and reuse publicly available components across ground and flight software stacks~\cite{curbo2023research}.
Agency initiatives at NASA and ESA further support this trend, making shared open-source building blocks increasingly available for space software development~\cite{nasa2025cfs,esa_oss_resources}.

This adoption of OSS is growing across the sector, including onboard systems, where open-source frameworks are integrated into critical satellite software.
The primary goal of adopting OSS in the space sector is to accelerate development and lower the barrier to entry; by leveraging existing, community vetted components for tasks like Command and Data Handling or payload management, teams can avoid developing every component from scratch~\cite{cratere2024board,thibault2022spaceflight}.
This approach, which prioritizes reuse and rapid prototyping, is particularly prevalent in academic and commercial programs.
In some cases, entire flight-software repositories are made publicly available to support collaboration, reuse, and standardization~\cite{curbo2023research}.

\subsection{Software Security Analysis and Supply Chain Exposure}

Automated software security analysis is used to identify weaknesses in both project developed code and third party components.
SAST examines source code for insecure programming patterns, including memory-safety errors, improper input validation, unsafe API use, and other implementation-level weaknesses.
IaC analysis and secret scanning complement this process by identifying insecure configurations and exposed credentials.
Together, these techniques provide a code-level perspective that complements dependency-focused analysis~\cite{lin2024vulnerabilities}.

SCA tools are widely used to identify, analyze, and manage open-source and third-party components in a codebase.
Modern applications are often assembled from many external packages drawn from ecosystems such as PyPI, Debian, and GitHub.
Manually assessing each component and its direct and transitive dependencies against the thousands of vulnerabilities disclosed each year is impractical.
For this reason, SCA tools are commonly used as an automated first step in managing such risks and helping secure the software supply chain~\cite{chahar2012code}.

These tools typically build or infer a dependency inventory, often supported by SBOM generation, and cross-reference identified components with public vulnerability databases such as the National Vulnerability Database (NVD).
This process identifies known flaws cataloged as CVEs and helps determine whether a project depends on vulnerable package versions.
Such analysis is important because adversaries actively target OSS ecosystems, package repositories, and update channels~\cite{ladisa2024understanding,ohm2020backstabber}.
A single high impact vulnerability in a widely used open-source component can propagate across many downstream projects.
Log4Shell is a prominent example of such cascading exposure, with impacts reported in mission adjacent aerospace systems~\cite{9857849}.

Recent work synthesizes methods for vulnerability detection and security patch detection in OSS and outlines a vulnerability-to-patch lifecycle~\cite{lin2024vulnerabilities}.
It emphasizes that managing software security exposure requires identifying weaknesses in project developed code, detecting vulnerable dependencies, and verifying patch application and remediation status.
This supports combining source code analysis, dependency analysis, and explicit patch status checks when assessing risk.

\begin{figure*}[htbp]
  \centering
  \includegraphics[width=\textwidth]{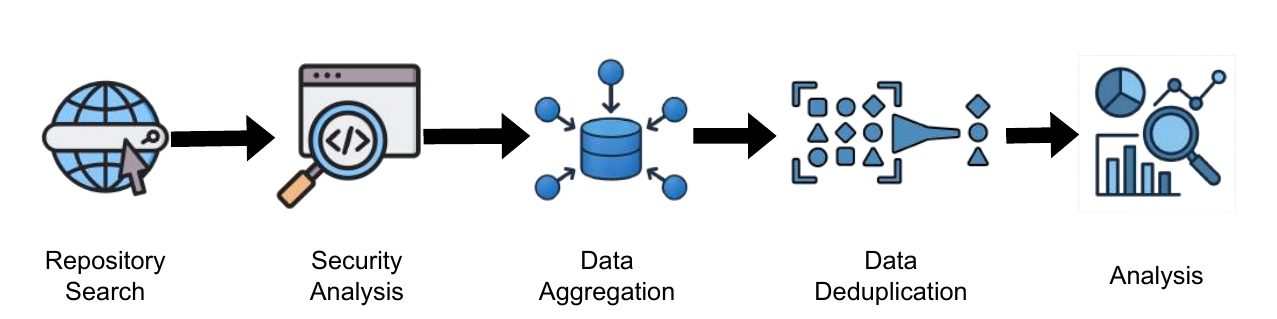}
  \caption{Multi-stage analysis pipeline: collection, aggregation, and analysis.}
  \label{fig:pipeline}
\end{figure*}

\subsection{Empirical Security Analysis of Open-Source Software}

Prior work has applied automated security analysis to large collections of OSS repositories to characterize recurring weaknesses and assess the usefulness of different analysis approaches.
Hashmat et al.~\cite{hashmat2024insights} applied 24 static analysis tools to thousands of OSS repositories, while Opri{\c{s}}a et al.~\cite{opricsa2026large} used a multi-tool framework to analyze more than ten thousand public repositories.
Both studies demonstrate the value of combining multiple analysis tools to obtain a broader view of software-security issues and highlight the importance of aggregating and interpreting findings produced by different tools.
Aloraini et al.~\cite{aloraini2019empirical} similarly showed that different SAST tools emphasize different types of weaknesses, with input-validation, API-related, and code-quality findings recurring across the analyzed C++ projects.

The validity and coverage of automated findings have also been examined.
Lipp et al.~\cite{lipp2022empirical} evaluated six static analyzers against real-world OSS with known vulnerabilities and found that individual tools missed a substantial portion of the vulnerabilities.
Combining multiple analyzers improved detection by 21--34 percentage points, although the resulting combinations still missed between 30\% and 69\% of the known vulnerabilities.
These findings reinforce the value of multi-tool analysis while also showing that automated results should be interpreted as security findings rather than complete ground truth.
Other empirical studies have examined security from different perspectives.
Al-Shammare et al.~\cite{al2024empirical} analyzed recurring CWE weaknesses across C\#, C++, and Java OSS projects, while Zahedi et al.~\cite{zahedi2018empirical} examined security-related issues reported by developers in GitHub repositories.

Research on open-source embedded software is particularly relevant to onboard systems.
Shen et al.~\cite{shen2025finding} applied CodeQL to 258 embedded-software projects and identified recurring security-relevant defects, including null-pointer handling, allocation-size errors, unbounded writes, and other memory-related weaknesses.
Their study also showed that embedded software may require domain-specific consideration of project structure, build systems, and analysis configuration when applying general-purpose security tools.

Our work builds on this body of empirical OSS security research but focuses specifically on publicly available software used in or developed for onboard satellite systems.
Rather than studying general OSS or evaluating individual analysis tools, our goal is to identify recurring security patterns within OSS for onboard satellite software domain and examine where these weaknesses occur.
This domain-specific perspective is intended to provide insight into the security characteristics of onboard OSS and support the development, integration, and maintenance of more secure satellite software.
\section{Methodology}
\label{sec:methodology}

To conduct our empirical study of publicly available OSS relevant to onboard satellite systems, we designed an analysis pipeline.
The pipeline consists of three stages: repository collection and scope definition, security analysis using a diverse set of open-source tools, and data aggregation into a consolidated dataset.
Figure~\ref{fig:pipeline} provides an overview of the pipeline.
We additionally performed validation reviews to assess the reliability of the resulting dataset.

\subsection{Repository Collection and Scope}

Our pipeline begins with the curation of a dataset of open-source repositories relevant to onboard satellite systems.
We identified candidate repositories from public code-hosting platforms, primarily GitHub and GitLab, using satellite- and flight-software-related search terms such as ``satellite,'' ``LEO,'' ``CubeSat,'' ``flight software,'' and ``cFS.''
This search produced a broad set of space-related projects, including flight-software repositories, reusable onboard frameworks, payload software, subsystem software, and supporting libraries.

The candidate repositories were manually assessed by two researchers to determine whether they contained software used in onboard satellite systems or software developed for future integration and operation onboard satellites.
Repositories containing flight software, onboard frameworks, subsystem software, payload software, or supporting libraries for onboard systems were retained.
The review excluded ground-segment-only tools, documentation-only repositories, simulation-only projects, unrelated forks, and projects outside the satellite-software scope.
For repositories that were retained, we collected metadata including creation date and last update.
This process resulted in a final dataset of 126 public repositories for analysis, which are listed in ~\ref{app:git-repos}.

\subsection{Security Analysis Toolchain}

Rather than relying on a single scanner, our pipeline applies a diverse set of open-source security analysis techniques, including SBOM generation, SCA, SAST, IaC checks, and secret scanning.
This multi-tool approach allows us to observe different classes of security-relevant findings, including vulnerable dependencies, code-level weakness patterns, configuration issues, and exposed secrets.
For each of the 126 repositories, we cloned the default branch at the time of scanning and applied the tool suite subject to tool applicability, language support, and available project files.
Tool outputs were collected in their native report formats and parsed into a common normalized schema for aggregation.

\begin{itemize}
    \item \textbf{SCA:}
    To identify known vulnerabilities in third-party components, we used a combination of SCA tools.
    We used Syft~\cite{syft} to generate a CycloneDX SBOM, followed by Grype~\cite{grype} for vulnerability analysis.
    We also ran Trivy~\cite{trivy}, osv-scanner~\cite{osv_scanner}, and the OWASP dependency-check tool~\cite{dependencycheck} to detect known vulnerable packages and dependency exposure.

    \item \textbf{SAST:}
    To identify insecure coding patterns, we used Semgrep~\cite{semgrep} with its default ``auto'' ruleset, Bearer~\cite{bearer} for general code-security risks, CodeQL~\cite{codeql} for semantic code analysis, and language-specific analyzers.
    For Python projects, Bandit~\cite{bandit} was used.
    For C/C++ projects, cppcheck~\cite{cppcheck} was run with all checks enabled.
    Language-specific tools were skipped when no relevant source files were present in the repository.

    \item \textbf{IaC \& misconfiguration:}
    To detect misconfigurations in deployment and build-related files, we used Checkov~\cite{checkov}.
    The scan included files such as Dockerfile, YAML, and Terraform (.tf) files when present.

    \item \textbf{Secret scanning:}
    To identify hardcoded credentials, API keys, and other sensitive values, we used Gitleaks~\cite{gitleaks}.
\end{itemize}

Tool execution status was recorded for each repository to distinguish completed scans from tools that were not applicable based on language support or available project files.
A repository was classified as having no detected findings only when its applicable analyses completed without producing onboard-relevant findings.
The versions of the security analysis tools and their scan configurations are provided in ~\ref{app:tool-config}.

\subsection{Data Aggregation, Filtering, and Deduplication}

The raw outputs from the toolchain were processed into a single normalized dataset.
Custom parsers ingested the native reports produced by each tool and mapped them into a common schema containing the repository, tool, severity, finding title, weakness identifier, dependency information when available, source location when available, and supporting metadata.
Repository identifiers and file paths were normalized to support consistent filtering and comparison across tools.
Severity values were normalized into four levels.
When available, CVSS-based scores associated with CVEs or vulnerability advisories were used, followed by tool-defined numeric security scores when such scores were unavailable.
When numerical severity information was unavailable, categorical severity labels were mapped to the four-tier scale according to each tool's native severity model.

We then applied a staged filtering process.
First, rule-based cleaning removed non-actionable findings and recurring tool-specific artifacts, including informational results, generated artifacts, repository metadata, automation paths, and other findings identified through documented cleaning rules.
Second, onboard-scope filtering removed findings associated with ground-segment-only code, documentation, examples, training artifacts, simulator-only paths, and other locations outside the onboard software scope.
This staged process preserved an audit trail of removed findings while focusing the final dataset on security-relevant findings in onboard-relevant code and dependencies.
Ambiguous scope decisions were subsequently reviewed through version-controlled manual restoration and removal rules, with the supporting rationale preserved in the audit trail.

We deduplicated the remaining findings using tier-specific fingerprints.
For SCA findings, duplicates were identified using repository, package name, package version, and vulnerability identifier.
For SAST, IaC, and secret-scanning findings, duplicates were identified using repository, source location, rule or weakness identifier, and available contextual metadata.
When multiple tools reported the same underlying finding, we retained the highest severity while preserving metadata about all tools that reported it.

Table~\ref{tab:tool_execution_coverage} summarizes the repository-level execution coverage of each analysis tool and its contribution to the final deduplicated dataset.
Language- and project-specific tools were classified as not applicable when the repository did not contain the supported source language or relevant project files.

\begin{table}[t]
\centering
\scriptsize
\setlength{\tabcolsep}{3pt}
\renewcommand{\arraystretch}{1.05}
\caption{Repository-level execution coverage and contribution to the final deduplicated dataset by analysis tool.}
\label{tab:tool_execution_coverage}
\begin{tabular}{lrrr}
\toprule
\textbf{Tool} &
\shortstack{\textbf{Successful}\\\textbf{Scans}} &
\shortstack{\textbf{Not}\\\textbf{Applicable}} &
\shortstack{\textbf{Retained}\\\textbf{Findings}} \\
\midrule
\texttt{Syft}             & 126 & 0   & 0     \\
\texttt{Grype}            & 126 & 0   & 87    \\
\texttt{Trivy}            & 126 & 0   & 36    \\
\texttt{Semgrep}          & 126 & 0   & 122   \\
\texttt{Gitleaks}         & 126 & 0   & 35    \\
\texttt{osv-scanner}      & 126 & 0   & 20    \\
\texttt{dependency-check} & 126 & 0   & 130   \\
\texttt{Bandit}           & 66  & 60  & 283   \\
\texttt{Checkov}          & 73  & 53  & 23    \\
\texttt{Bearer}           & 76  & 50  & 284   \\
\texttt{cppcheck}         & 113 & 13  & 1,483 \\
\texttt{CodeQL}           & 126 & 0   & 324   \\
\midrule
\multicolumn{3}{l}{\textbf{Total retained findings}} &
\textbf{2,827} \\
\bottomrule
\end{tabular}
\end{table}

To better characterize the findings, we used the OWASP Top 10:2025 to group them into broad security-risk categories and CWE to identify the relevant software weakness types.
The OWASP Top 10 is a widely recognized industry baseline for application security risks and provides a useful means of categorizing our results~\cite{owasp_top10_2025}.
CWE is a community-developed taxonomy of software and hardware weaknesses that provides a more granular description of the specific weakness types represented in the findings~\cite{mitre_cwe_2026}.
To apply the CWE taxonomy consistently, each finding was assigned to one of eight higher-level weakness families through a structured mapping process.
Findings with a reported CWE were mapped directly using the defined CWE-family taxonomy, while findings without a reported CWE were classified using the finding title, description, scanner classification, and available code context.
Each finding was then assigned to the most relevant single OWASP Top 10:2025 category, prioritizing official CWE associations and, where no direct association was available, using documented mappings derived from the finding title, description, scanner classification, and available code context.

The findings were also classified by code source and onboard subsystem.
The code-source classification identifies whether a finding occurs in project-developed code or an external dependency.
The subsystem classification assigns each finding to the relevant onboard subsystem, such as the onboard computer (OBC), payload, telemetry, tracking, and command (TTC), attitude determination and control system (ADCS), electrical power system (EPS), or thermal control.
Assignments were based on repository structure, file paths, build configurations, source code, and project documentation.
Findings that supported multiple subsystems, or for which no single subsystem could be identified, were assigned to Shared Runtime / Cross-Subsystem.

The findings were further mapped using SPARTA (Space Attack Research and Tactic Analysis), a framework for analyzing cyber threats and adversarial techniques in space systems~\cite{slay2023applying}.
We used SPARTA's Threats to Space Systems library, which defines threat categories for spacecraft systems~\cite{sparta_threats_space}.
Each finding was assigned to a single threat category based on the identified weakness mechanism and the available onboard software context, including the affected component, source-code location, subsystem assignment, and supporting finding metadata.

To assess onboard software update capability, we manually reviewed all 126 repositories using available source code and project documentation.
An update capability was identified when onboard software could be replaced or updated after deployment, either directly through the repository or through a parent platform or framework.
The classifications were independently validated, including checks for framework-level update paths, using evidence available in the current public repository snapshots.

\subsection{Onboard-Scope and Finding Validation}

To assess the accuracy and coverage of the onboard-scope classification, we reviewed all findings in the final dataset to verify that they were associated with onboard-relevant software.
Each finding was assessed using the available evidence, including the reporting tool, severity, weakness identifier, dependency information, source location, code context when available, and supporting metadata.
In addition, a separate sample of excluded findings was reviewed to assess whether onboard-relevant findings had been incorrectly removed during filtering.
This validation focused on the accuracy of the onboard-scope classification and on potential omissions introduced during filtering.

Separately, to assess scanner accuracy in the retained dataset, we reviewed a random sample of 360 unique findings.
For each sampled finding, we examined the corresponding scanner output and, where available, the source-code context, dependency and advisory information, and relevant configuration.
Findings were classified according to whether the available technical evidence supported or contradicted the condition reported by the scanner.
Where additional context was required, the available source code and supporting evidence were further examined before assigning the final classification.

The goal of these validation steps is to assess the reliability of the retained dataset by verifying onboard relevance and estimating the false-positive rate of the retained findings.
The validation was not intended to establish mission-specific exploitability or runtime reachability.
Accordingly, this study characterizes the ecosystem-level distribution and recurring patterns of automated security findings across onboard-relevant OSS, with the validation analyses used to assess the reliability of the resulting dataset.
\section{Results}
\label{sec:results}

Our analysis pipeline was applied to 126 public repositories containing onboard-relevant satellite software.
The tools produced 28,274 candidate vulnerability findings, which were normalized to 18,894 raw parsed findings and then processed through false-positive cleaning, onboard-scope filtering, and deduplication to obtain a final dataset of 2,827 candidate vulnerability findings.
This section presents the resulting security landscape at both the dataset and repository levels.
We first describe how the raw scanner outputs were reduced to the final dataset and summarize the overall severity distribution.
Next, we examine how findings are distributed across repositories, assess onboard software update capability, and examine onboard software subsystems and code sources, before characterizing the most common weakness categories using the OWASP Top 10 and CWE, and examining their mapping to SPARTA Threats to Space Systems.

\subsection{Dataset Construction}

\begin{figure}[t]
    \centering
    \includegraphics[width=\linewidth]{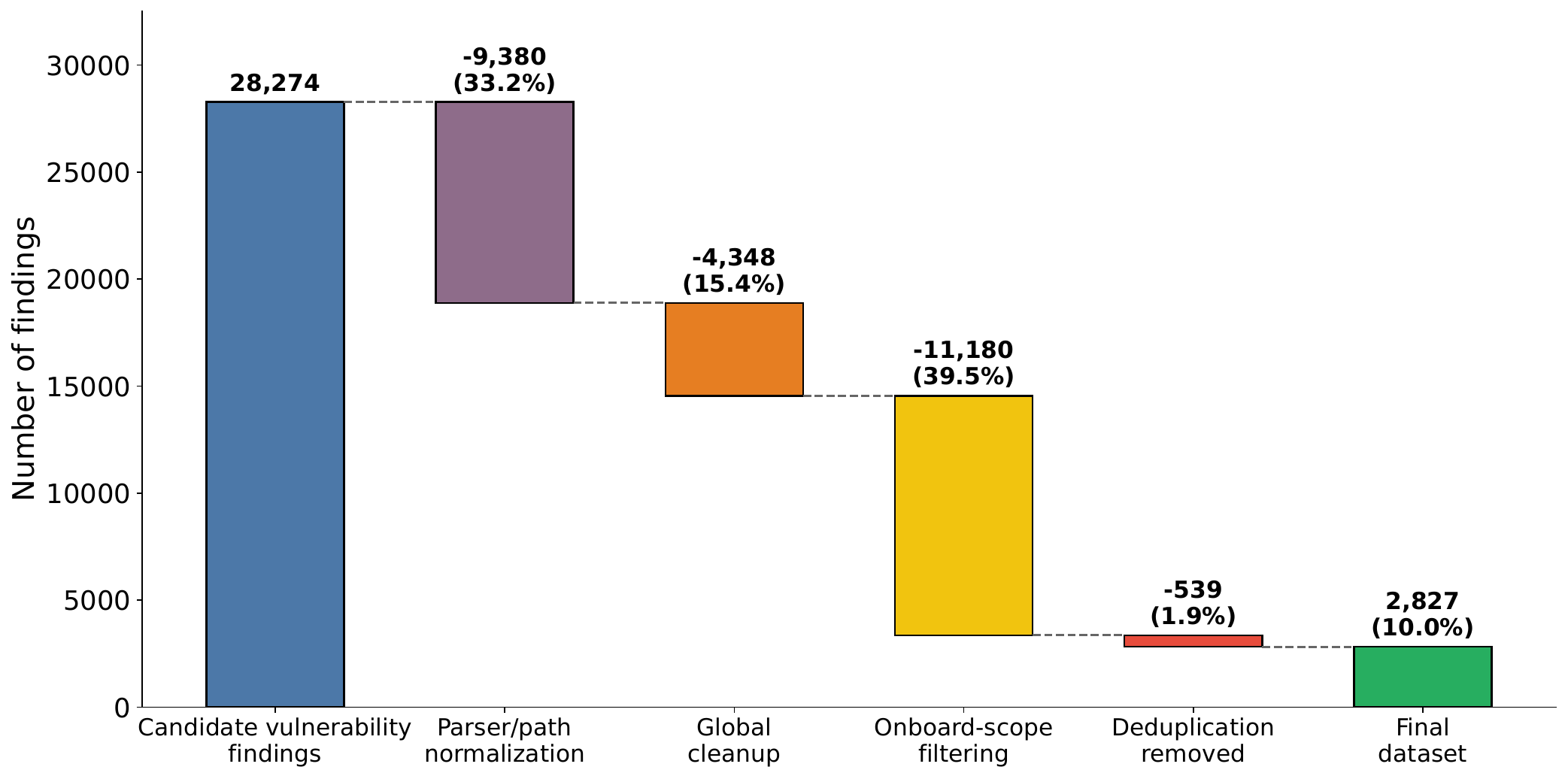}
    \caption{Dataset construction from candidate vulnerability findings to the final analysis dataset after parser/path normalization, global cleanup, onboard-scope filtering, and deduplication. Percentages shown in the figure are calculated relative to the initial 28,274 candidate vulnerability findings.}
    \label{fig:dedup_waterfall}
\end{figure}

As shown in Figure~\ref{fig:dedup_waterfall}, the scanner outputs contained 28,274 candidate vulnerability findings across the repository corpus.
Parser and path normalization first reduced these outputs to 18,894 raw parsed findings by excluding non-admissible records such as documentation, tests, examples, vendor paths, ground-support paths, and tool records outside the analysis scope.
We then processed the raw parsed findings through three stages to construct the final analysis dataset.
First, global cleanup removed 4,348 findings, corresponding to 23.0\% of the 18,894 raw parsed findings.
This stage removed non-actionable findings and recurring scanner artifacts, including informational reports, generated files, repository metadata, automation paths, and known tool-specific false-positive patterns.
Second, onboard-scope filtering removed an additional 11,180 findings by excluding findings associated with ground-only code, documentation, examples, simulators, build artifacts, and other paths not treated as onboard-relevant software.
A breakdown of the filtering reasons is provided in ~\ref{app:filtering-breakdown}.
Finally, fingerprint-based deduplication removed an additional 539 findings, producing the final dataset of 2,827 candidate vulnerability findings shown in Figure~\ref{fig:dedup_waterfall}.

Following the scope-review corrections, all 2,827 findings in the final dataset were reviewed, and no remaining out-of-scope findings were identified.
A separate sample of 150 excluded findings was also reviewed, and no onboard-relevant findings were found to have been incorrectly removed during filtering.

\subsection{Overall Vulnerability Landscape}

In our analysis, we adopted a standard four-tier severity model (low, medium, high, critical) to classify the candidate vulnerability findings.
Because the tools use different native severity representations, the normalized levels provide a common triage scale for cross-tool aggregation and comparison.
These severity levels reflect the potential impact reported by the analysis tools and support the prioritization of findings, although actual exploitability depends on the software configuration, runtime reachability, and operational context.
Critical findings represent potentially severe risks that warrant immediate attention, such as hardcoded administrative credentials.
High severity findings may enable privilege escalation, data exfiltration, or denial of service, while medium severity findings may become exploitable under specific conditions or in combination with other weaknesses, such as improper input handling or weak cryptography.
Finally, low severity findings generally represent best-practice violations or issues with low potential impact or high exploit complexity.

\begin{figure}[htbp]
  \centering
  \includegraphics[width=0.8\linewidth]{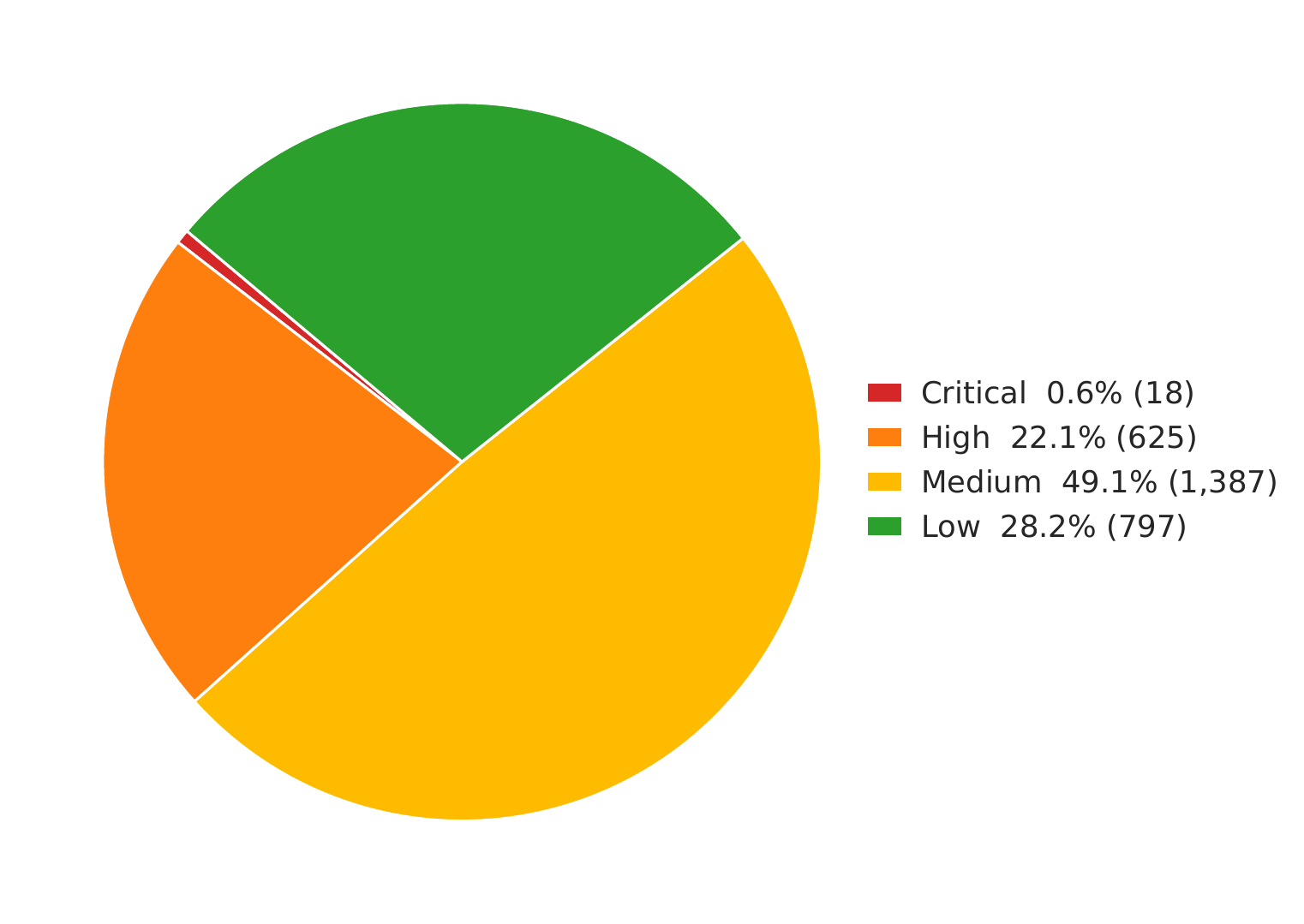}
  \caption{Distribution of the 2,827 candidate vulnerability findings by severity.}
  \label{fig:severity_all}
\end{figure}

Figure~\ref{fig:severity_all} shows the distribution of all findings by severity across the dataset.
In this paper, we treat medium or higher severity findings as priorities for further triage, while low severity findings are tracked as baseline hygiene.
Medium severity findings were the most common, accounting for 49.1\% of all findings (1,387).
This was followed by low severity findings, which accounted for 28.2\% (797), high severity findings, which accounted for 22.1\% (625), and critical findings, which accounted for 0.6\% (18).
Overall, medium and higher severity findings account for 71.8\% of all findings (2,030), while high and critical findings together account for 22.7\% (643).
The high prevalence of medium and higher severity findings shows that a large share of the identified weaknesses warrants review and prioritization.

\subsection{Distribution Across Repositories}

\begin{figure}[htbp]
\centering
\includegraphics[width=\linewidth]{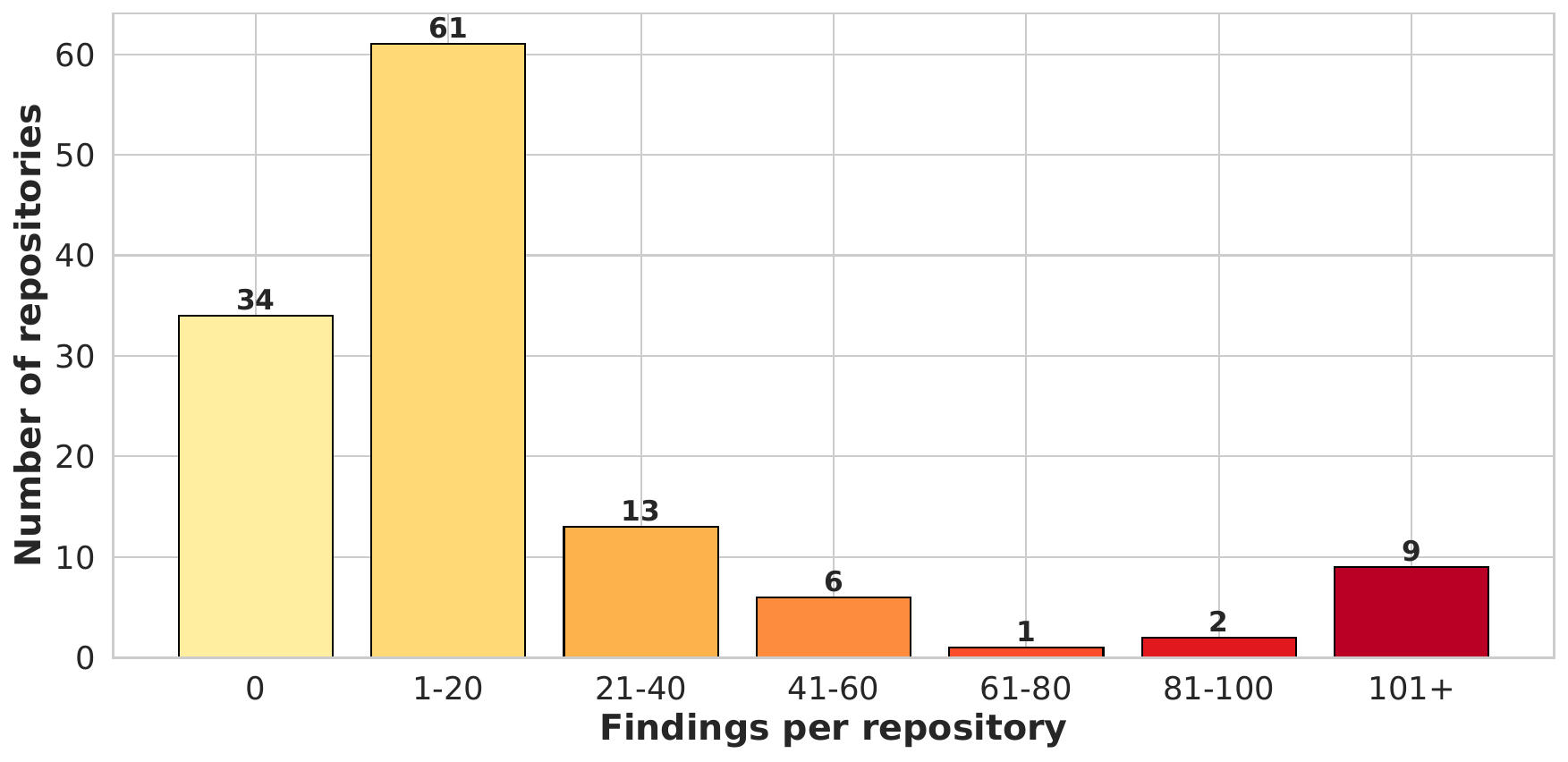}
\caption{Distribution of repositories by total finding count across all 126 repositories.}
\label{fig:repo-dist-all}
\end{figure}

We next examine how the findings are distributed across the repositories.
A key result of our study is that the findings are not distributed evenly across the ecosystem.
Figure~\ref{fig:repo-dist-all} shows the distribution of repositories by total finding count.
Thirty-four repositories (27.0\%) had no findings in the final dataset, and 61 repositories (48.4\%) had 1--20 findings.
Repositories with no retained findings should not be interpreted as secure, as our study provides a high-level static analysis of onboard-relevant software rather than a comprehensive security assessment of each repository.
Together, these groups account for 75.4\% of the corpus.
At the other end of the distribution, nine repositories contained more than 100 findings, showing that a substantial portion of the dataset is concentrated in a small number of repositories.

Repository metadata showed that 59 of the 126 repositories (46.8\%) had not been updated for more than two years at the time of analysis.
To examine whether repository maintenance status was associated with the observed findings, we compared these 59 stale repositories with the 67 repositories updated within the previous two years.
Stale repositories contained an average of 28.98 findings per repository, compared with 16.67 among maintained repositories.
They also contained an average of 7.42 High or Critical findings per repository, compared with 3.06 among maintained repositories.

\begin{figure}[htbp]
  \centering
  \includegraphics[width=\linewidth]{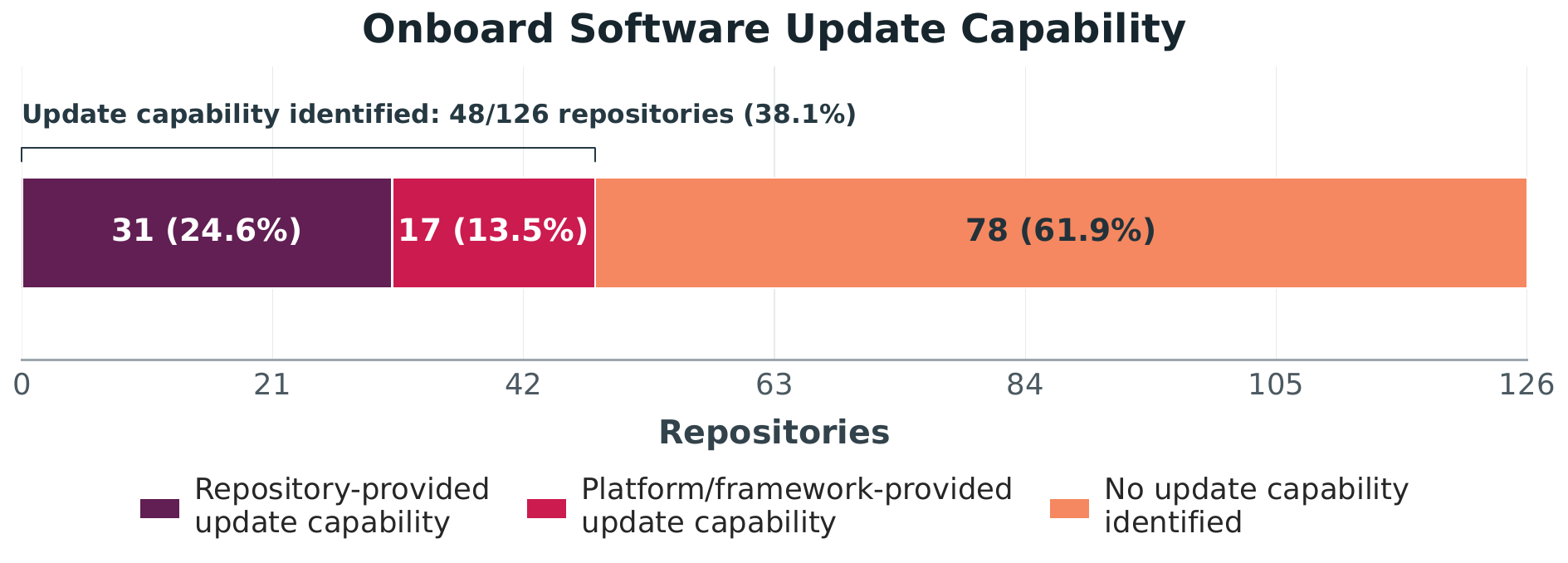}
  \caption{Onboard software update capability across the 126 analyzed repositories.}
  \label{fig:update_capability}
\end{figure}

We also examined onboard software update capability across the repository corpus.
As shown in Figure~\ref{fig:update_capability}, an update capability was identified for 48 of the 126 repositories (38.1\%).
Of these, 31 repositories (24.6\%) provided an update capability directly, while 17 additional repositories (13.5\%) relied on a documented update mechanism provided by their parent platform or framework.
No update capability was identified for the remaining 78 repositories (61.9\%), indicating that onboard software update support is not widespread across the analyzed corpus.

\subsection{Distribution Across Subsystems and Code Sources}

To examine where the findings occur within onboard software, we classified them by subsystem and code source.
This subsystem-level view shows which parts of the satellite contain the largest number of findings.
The code-source classification shows whether the findings come from project-developed code or external dependencies.

\begin{figure}[htbp]
  \centering
  \includegraphics[width=1.0\linewidth]{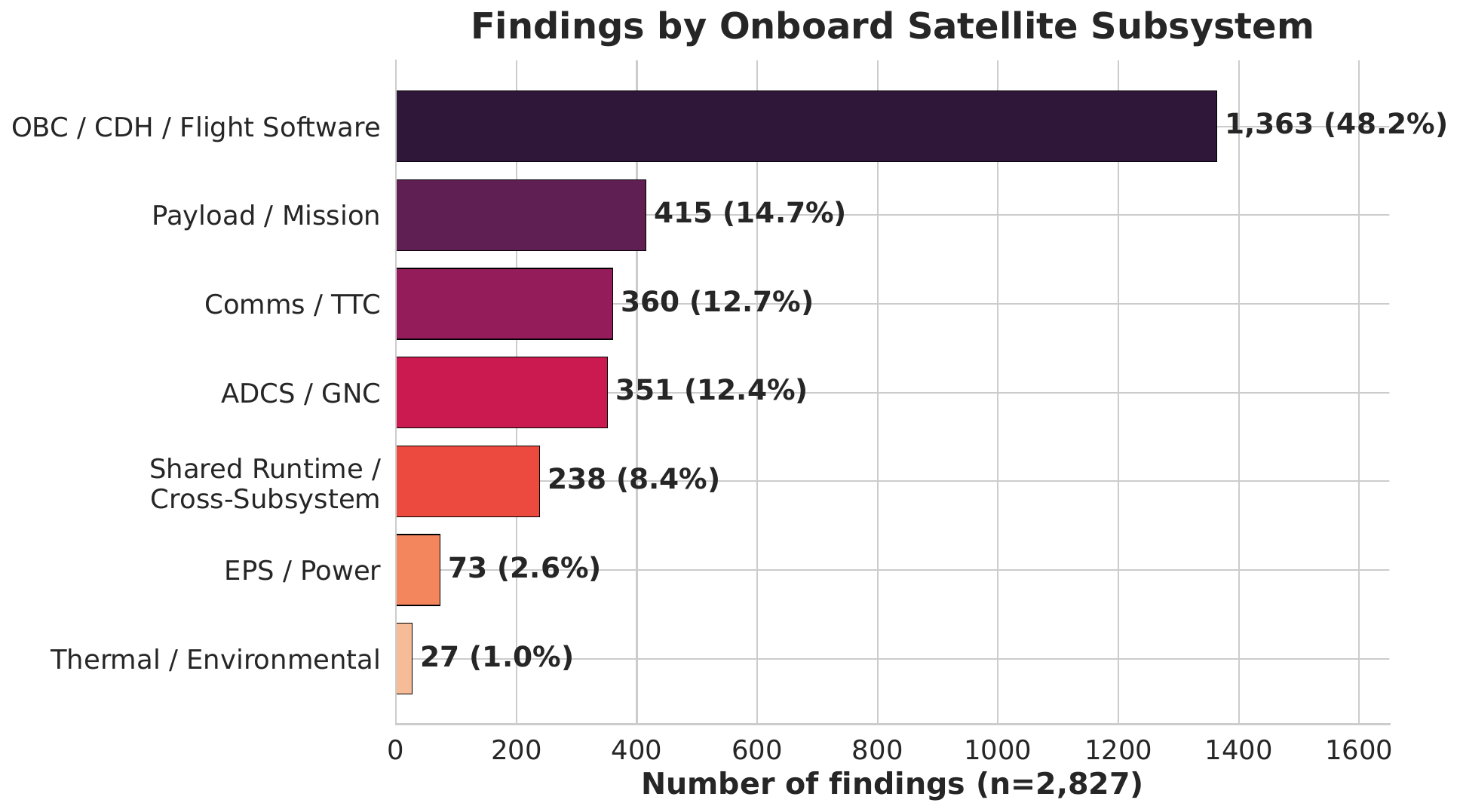}
  \caption{Distribution of the 2,827 findings across onboard software subsystems.}
  \label{fig:subsystem_distribution}
\end{figure}

Figure~\ref{fig:subsystem_distribution} shows the distribution of findings across the onboard subsystem categories.
OBC, CDH, and flight software contain the largest number of findings, accounting for 1,363 findings (48.2\%).
Payload and mission software account for 415 findings (14.7\%), followed by communications and TTC software with 360 findings (12.7\%).
Together, these three categories account for 2,138 findings, or 75.6\% of the final dataset.
ADCS and GNC software account for 351 findings (12.4\%), while thermal and environmental software and EPS and power software account for 27 (1.0\%) and 73 (2.6\%), respectively.
The remaining 238 findings (8.4\%) occur in shared runtime or cross-subsystem code that could not be assigned to a single subsystem.
It is important to note that these results show how findings are distributed across subsystems, not how vulnerable each subsystem is.

\begin{figure}[htbp]
  \centering
  \includegraphics[width=1.0\linewidth]{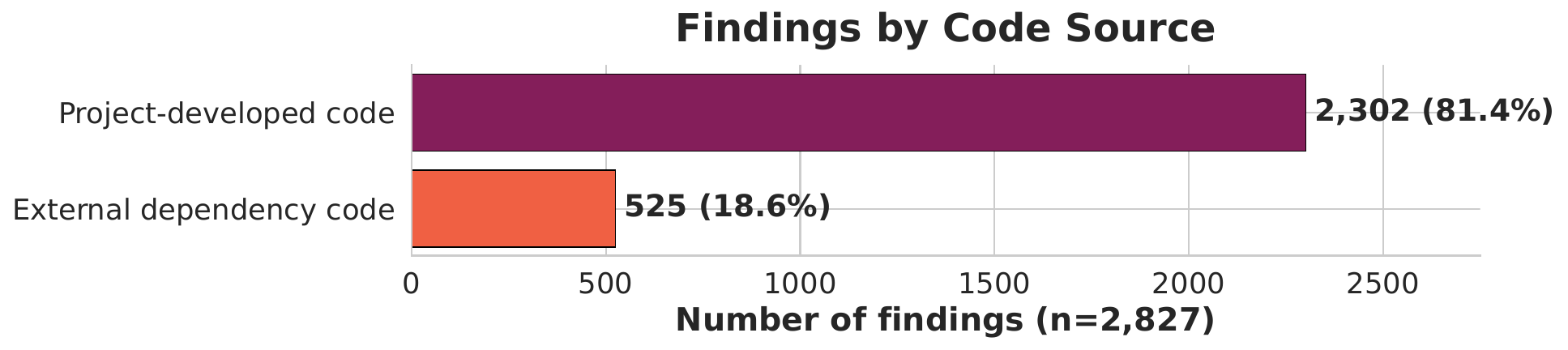}
  \caption{Distribution of the 2,827 findings between project-developed code and external dependency code.}
  \label{fig:code_source_distribution}
\end{figure}

Figure~\ref{fig:code_source_distribution} examines whether the findings are located in project-developed code or in external dependencies.
Project-developed code accounts for 2,302 findings (81.4\%), while external dependency code accounts for 525 findings (18.6\%).
The majority of findings therefore occur in code developed within the analyzed projects.
External dependencies still account for nearly one-fifth of the dataset, making them an important part of the observed security findings.
These results show that security review must consider both the project codebase and the dependencies it incorporates.

\subsection{Weakness Categories (OWASP Top 10 and CWE)}

\begin{figure}[htbp]
  \centering
  \includegraphics[width=1.0\linewidth]{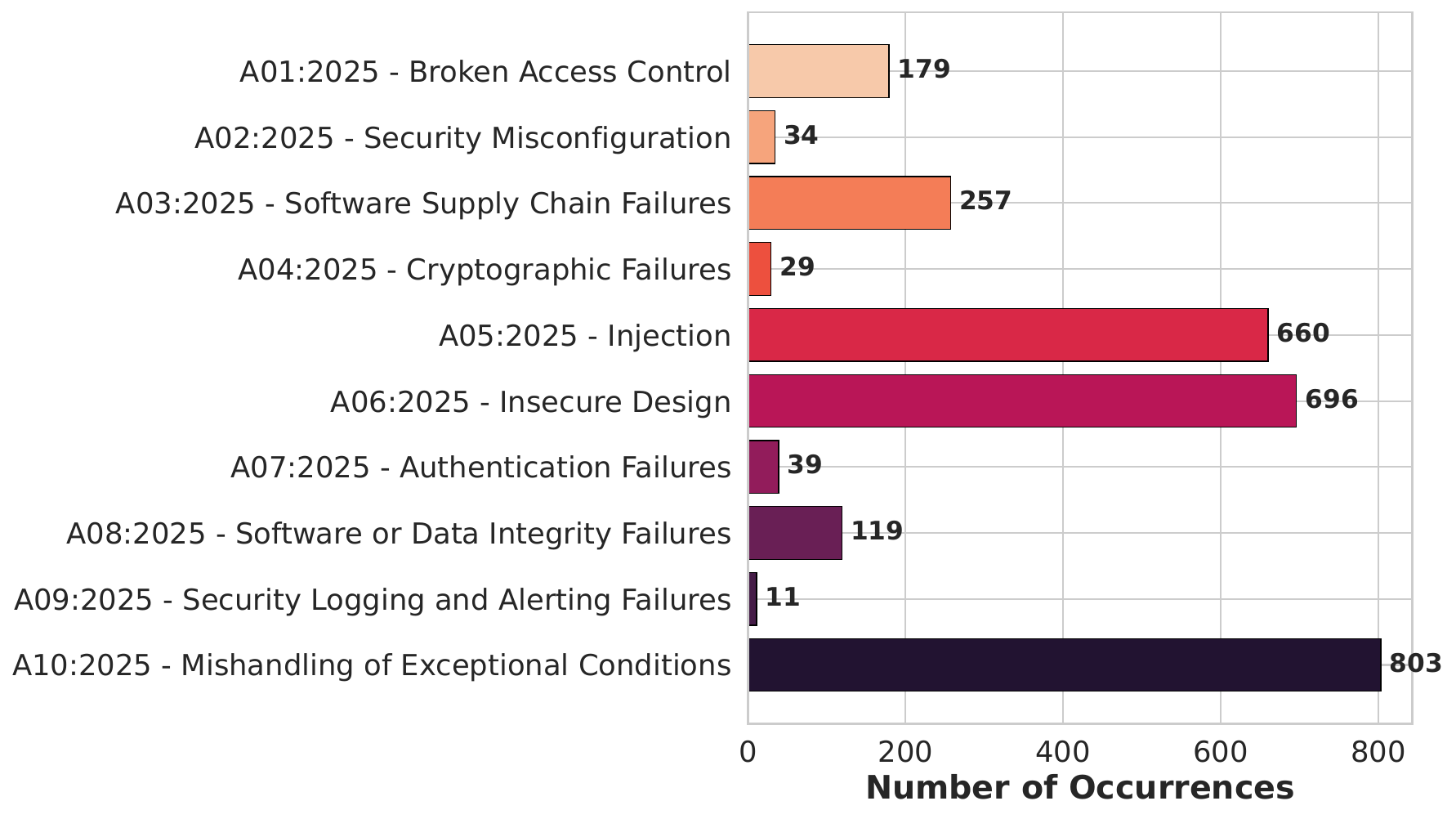}
  \caption{Distribution of the 2,827 findings across the OWASP Top 10:2025 categories.}
  \label{fig:owasp_top10}
\end{figure}

As shown in Figure~\ref{fig:owasp_top10}, each of the 2,827 findings was mapped to a single OWASP Top 10:2025 category.
In the aggregate dataset, Mishandling of Exceptional Conditions is the largest category, with 803 findings (28.4\%), followed by Insecure Design with 696 findings (24.6\%) and Injection with 660 findings (23.3\%).
At the repository level, Mishandling of Exceptional Conditions was also the most widespread category, appearing in 73 repositories, followed by Injection in 59 repositories and Insecure Design in 41 repositories.
The remaining categories account for smaller portions of the dataset, including Software Supply Chain Failures with 257 findings, Broken Access Control with 179, Software or Data Integrity Failures with 119, Authentication Failures with 39, Security Misconfiguration with 34, Cryptographic Failures with 29, and Security Logging and Alerting Failures with 11.

Depending on where they occur, Mishandling of Exceptional Conditions weaknesses may allow an attacker to trigger unhandled inputs or fault conditions that crash critical onboard services, disrupt dependent satellite functions, and, in severe cases, damage satellite components.
Insecure Design findings may expose weaknesses in trust boundaries or privilege separation that enable privilege escalation and increase the impact of an initial compromise.
Injection weaknesses may allow crafted inputs to reach a command-processing path and cause the execution of unintended commands on the satellite.

\begin{figure}[htbp]
  \centering
  \includegraphics[width=1.0\linewidth]{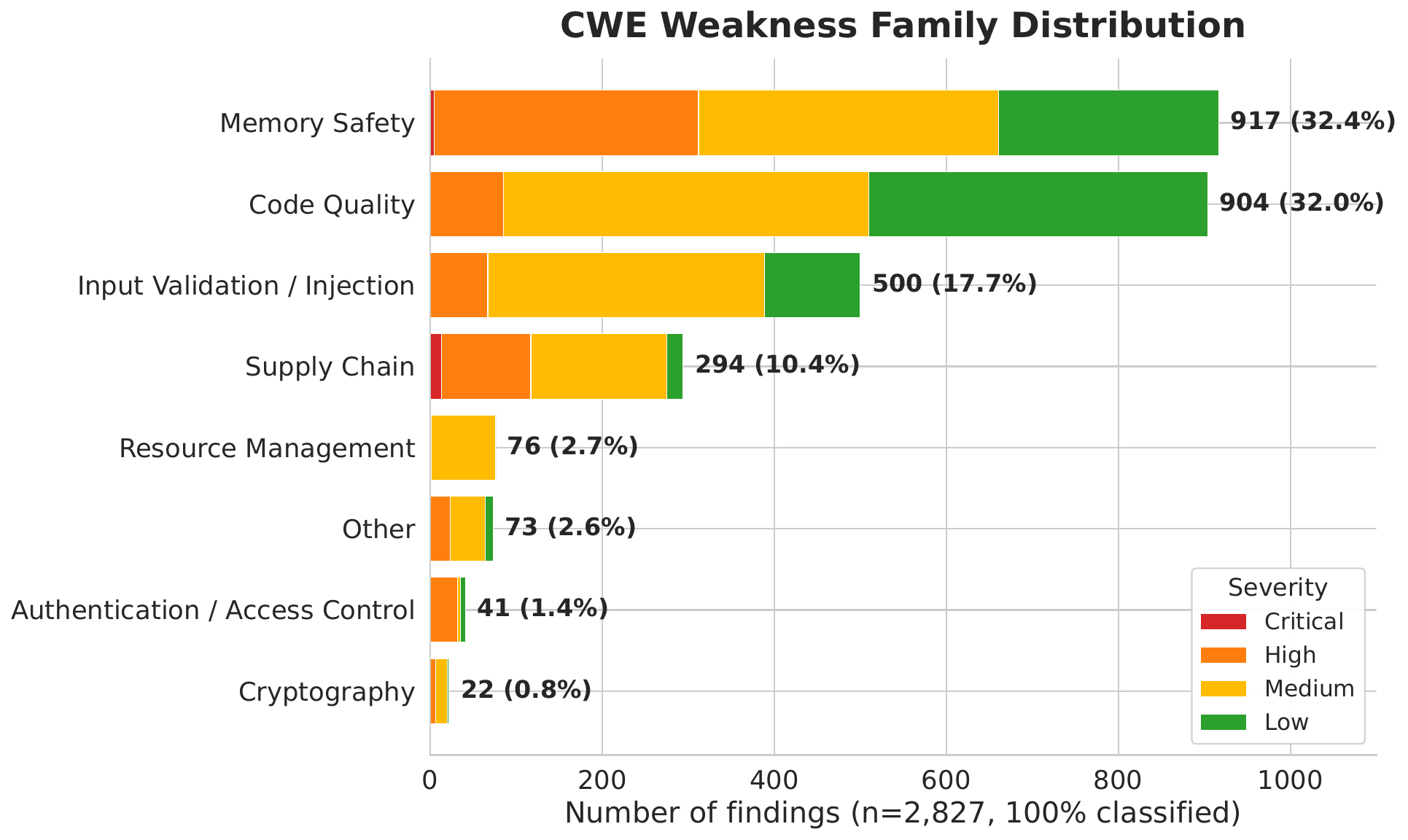}
  \caption{Distribution and severity composition of the 2,827 findings across the CWE weakness families.}
  \label{fig:cwe_family_distribution}
\end{figure}

Figure~\ref{fig:cwe_family_distribution} presents the distribution of the findings across eight CWE weakness families.
In the aggregate dataset, Memory Safety is the largest family, with 917 findings (32.4\%), followed by Code Quality with 904 findings (32.0\%) and Input Validation and Injection with 500 findings (17.7\%).
Together, Memory Safety and Code Quality account for 64.4\% of all findings.
The remaining families include Supply Chain with 294 findings (10.4\%), Resource Management with 76 findings (2.7\%), Other with 73 findings (2.6\%), Authentication and Access Control with 41 findings (1.4\%), and Cryptography with 22 findings (0.8\%).

Memory-safety weaknesses can allow attacker-controlled inputs to corrupt memory, crash onboard services, or potentially alter program execution.
Code-quality weaknesses can introduce undefined behavior or incorrect program logic that an attacker may use to trigger failures or amplify the impact of other vulnerabilities.
Input-validation and injection weaknesses can allow crafted inputs to reach sensitive operations and cause unintended actions within onboard software.

\begin{figure}[htbp]
  \centering
  \includegraphics[width=1.0\linewidth]{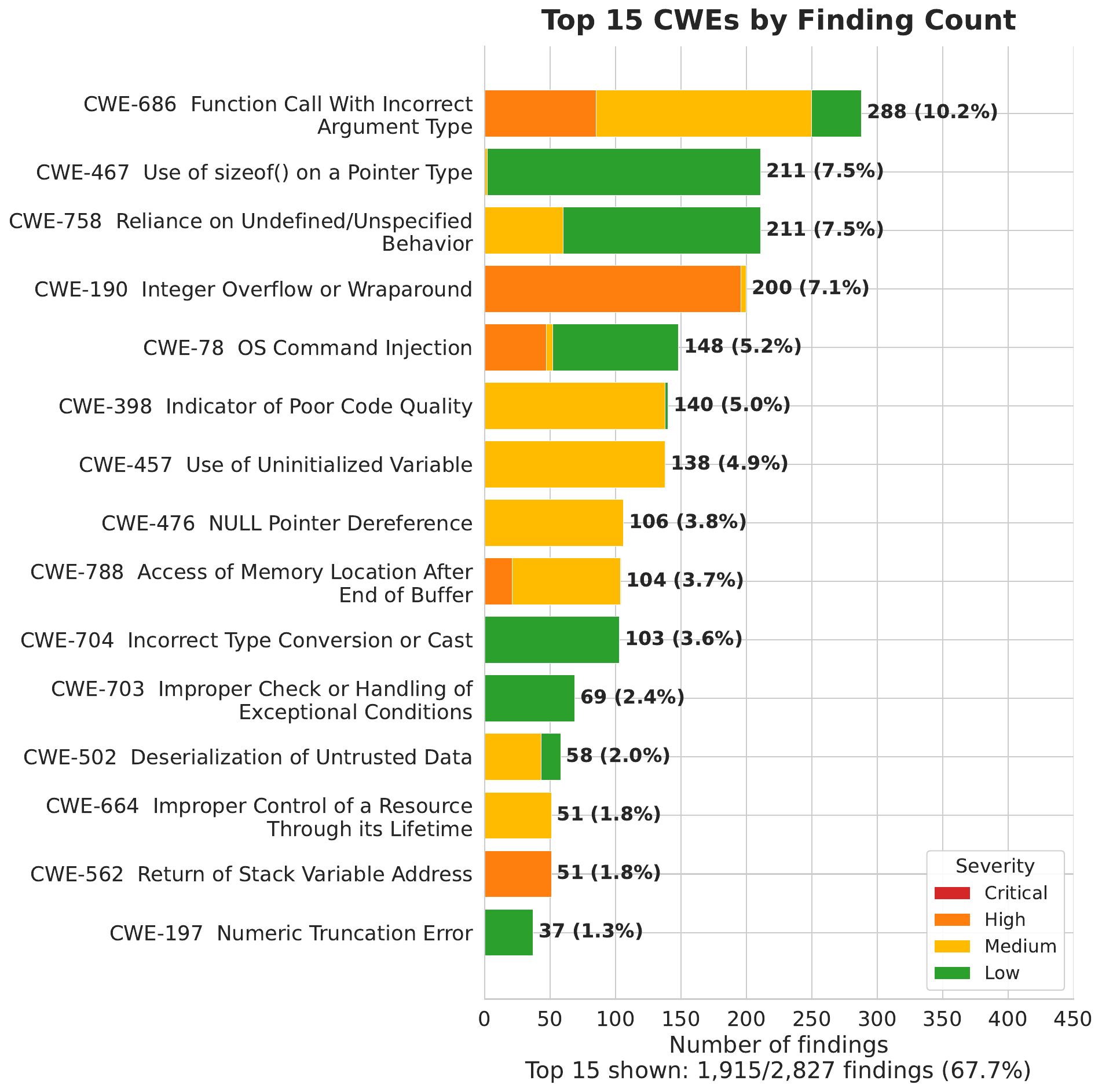}
  \caption{Fifteen most frequent CWE weakness types and their severity composition across the 2,827 findings.}
  \label{fig:top_cwes}
\end{figure}

Building on the family-level analysis, Figure~\ref{fig:top_cwes} examines the distribution at the level of individual CWEs and shows that the 15 most frequent CWEs account for 1,915 of the 2,827 findings (67.7\%).
CWE-686, Function Call With Incorrect Argument Type, is the most common with 288 findings (10.2\%), followed by CWE-467, Use of \texttt{sizeof()} on a Pointer Type, with 211 (7.5\%), and CWE-758, Reliance on Undefined or Unspecified Behavior, with 211 (7.5\%).
Other prominent CWEs include CWE-190, Integer Overflow or Wraparound, with 200 findings, and CWE-78, OS Command Injection, with 148.
The concentration of findings in these CWEs shows that incorrect function usage, pointer-size errors, undefined behavior, numeric errors, and command injection are recurring patterns across the analyzed onboard software.

\begin{figure}[htbp]
  \centering
  \includegraphics[width=1.0\linewidth]{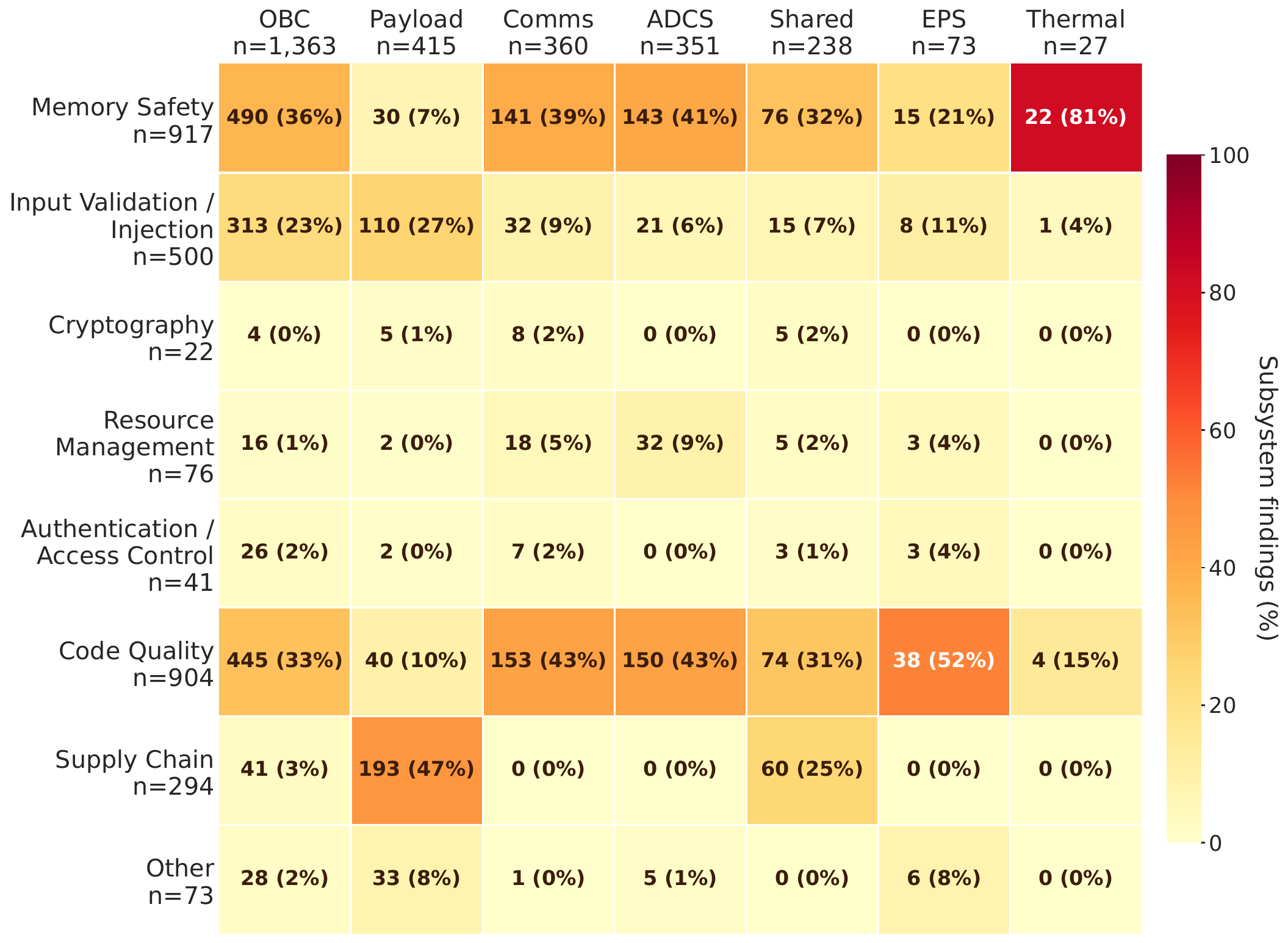}
  \caption{Distribution of findings across CWE families and onboard software subsystems. Each cell reports the number and percentage of findings assigned to a given CWE family within the corresponding subsystem.}
  \label{fig:cwe_subsystem}
\end{figure}

To examine whether these aggregate weakness patterns are consistent across onboard software functions, Figure~\ref{fig:cwe_subsystem} shows the CWE-family composition within each subsystem.
The distributions differ substantially across subsystem categories.
Memory Safety and Code Quality dominate OBC, communications, and ADCS software, jointly accounting for 68.6\%, 81.7\%, and 83.4\% of findings in these subsystems, respectively.
In contrast, Payload and mission software shows a distinct profile, with Supply Chain findings accounting for 46.5\% of its findings, compared with 3.0\% in OBC software and none in communications or ADCS.
EPS software is dominated by Code Quality findings (52.1\%), while Thermal and environmental software has a particularly high proportion of Memory Safety findings (81.5\%), although this category contains only 27 findings.

\subsection{SPARTA Threats to Space Systems}
\label{sec:sparta_threats}

\begin{figure}[htbp]
  \centering
  \includegraphics[width=1.0\linewidth]{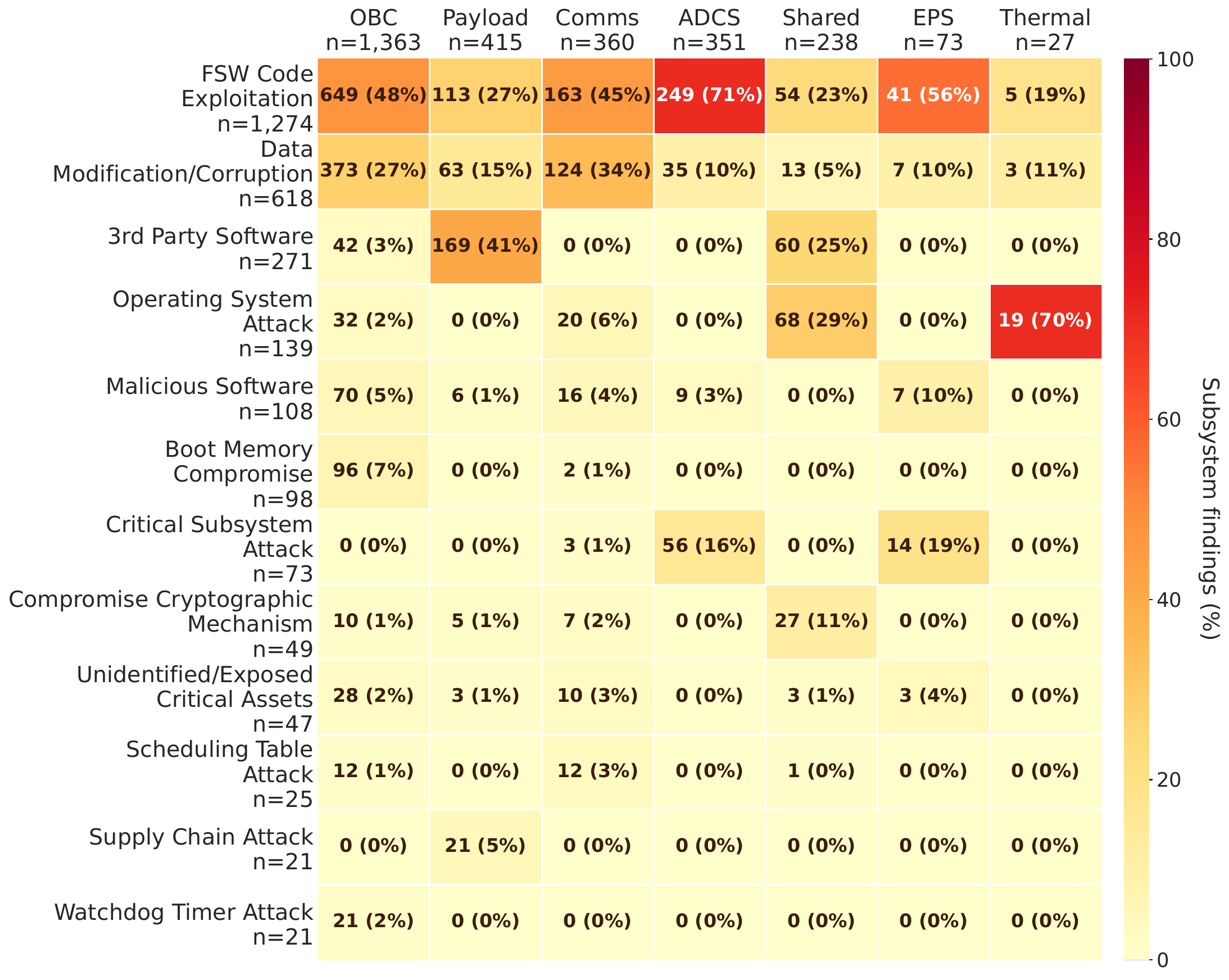}
  \caption{Distribution of findings across SPARTA Threats to Space Systems categories and onboard software subsystems for threat categories represented by at least 20 findings. Each nonzero cell reports the number and percentage of findings within the corresponding subsystem, calculated using all findings in that subsystem, including lower-frequency categories reported in ~\ref{app:sparta_threats}.}
  \label{fig:sparta_threats}
\end{figure}

Figure~\ref{fig:sparta_threats} presents the distribution of the findings across SPARTA Threats to Space Systems categories and onboard software subsystems.
The figure shows the 12 threat categories represented by at least 20 findings, which together account for 2,744 of the 2,827 retained findings.
Each nonzero cell reports the number of findings and their percentage within the corresponding subsystem.
The remaining nine lower-frequency threat categories are reported in ~\ref{app:sparta_threats}.

FSW Code Exploitation, which captures weaknesses in onboard flight software, is the most common threat category overall, accounting for 1,274 findings (45.1\%).
It is also the dominant category in OBC, communications, ADCS, and EPS software, accounting for 47.6\%, 45.3\%, 70.9\%, and 56.2\% of findings in these subsystems, respectively.
Data Modification/Corruption, which concerns unauthorized changes to spacecraft data, is also prominent in OBC (27.4\%) and communications software (34.4\%), while Critical Subsystem Attack, which captures threats affecting critical spacecraft functions, accounts for 16.0\% of ADCS findings and 19.2\% of EPS findings.
Payload and mission software shows a different profile, with 3rd Party Software, representing risks associated with reused, COTS, and open-source software, accounting for 40.7\% of its findings.
Shared runtime software is primarily associated with Operating System Attack (28.6\%) and 3rd Party Software (25.2\%), while Thermal and environmental software is dominated by Operating System Attack (70.4\%), although this subsystem contains only 27 findings.

\subsection{Validation and Sensitivity Analysis}
\label{sec:validation-sensitivity}

To estimate the false-positive rate of the retained findings, we reviewed a random sample of 360 unique findings from the retained dataset.
For each finding, we examined the corresponding scanner output and, where available, the source-code context, dependency and advisory information, and relevant configuration.
The review identified 22.2\% of the sampled findings as false positives (95\% CI: 17.6--27.6\%).
This rate is lower than those reported in some recent empirical evaluations, although direct comparison is limited by differences in tools, datasets, and validation methodologies~\cite{shen2025finding,opricsa2026large}.
This difference may partly reflect the stage at which validation is performed in our pipeline, as our sample is drawn from the retained dataset after rule-based cleaning, onboard-scope filtering, and deduplication.
Prior work similarly shows that filtering, scanner configuration, and analysis context can substantially affect false-positive rates~\cite{aloraini2019empirical,shen2025finding,opricsa2026large}.

To account for differences in evidentiary basis across finding types, we classified each retained finding into one of five finding classes: security-focused source-code alerts, correctness and quality warnings, vulnerable-dependency or advisory matches, secret candidates, and configuration or IaC findings.
We then analyzed the results by finding class and reporting tool.
Cppcheck accounts for 1,483 of the 2,827 retained findings, and sensitivity analysis shows that several aggregate trends depend strongly on its contribution.
Excluding Cppcheck reduces Memory Safety from 32.4\% to 17.3\% and Code Quality from 32.0\% to 13.9\%, while Injection becomes the largest OWASP category at 35.9\%.
In contrast, OBC, CDH, and flight software remain the largest subsystem category across the full dataset and the main sensitivity analyses.
These results distinguish tool-dependent category distributions from subsystem-level patterns that remain stable across different analysis views.

Severity was also analyzed by reporting tool and finding class because the tools use different native severity representations.
The normalized Critical, High, Medium, and Low categories provide a common triage scale for cross-tool aggregation and comparison.
The severity normalization procedure is provided in ~\ref{app:severity-mapping}.
\section{Discussion}
\label{sec:discussion}

Based on the results of the automated analysis pipeline, this discussion provides a high-level overview of the security findings observed across OSS used in onboard satellite systems.
The finding-validation analysis estimated a 22.2\% false-positive rate in the retained dataset.
While the identification of candidate vulnerability findings through automated tools highlights broader security risks, it is important to note that their actual exploitability on an operational satellite is uncertain.
Factors such as system configuration, runtime reachability, and the presence of an accessible attack surface, which determine if a vulnerability can truly be exploited, were not assessed in this study.
The goal of our analysis is therefore to identify recurring security patterns across a broad corpus of onboard-related OSS rather than to establish the runtime exploitability of individual findings.
Doing so would require executing the analyzed software within representative satellite configurations and operational environments, which is not feasible consistently across all repositories in the dataset.
For this reason, runtime techniques such as fuzzing and penetration testing were not conducted as part of this study.
Such techniques could complement our analysis by evaluating whether identified weaknesses, particularly runtime-dependent issues such as memory-safety defects, can be reached or triggered during execution.

Interpreting these results also requires considering the scope of the analyzed corpus.
Our dataset focuses on publicly available OSS repositories, including code that has been used in space as well as code intended to be integrated into mission builds.
Even when a repository represents only a subsystem or reference implementation that may be integrated and modified before deployment, the dependency choices and recurring coding patterns we observe can carry forward.
As a result, the same or similar classes of weaknesses may persist in operational software stacks.
Closed-source flight software used by commercial and government operators is not available for comparable analysis, so the recurrence and distribution of weaknesses observed in this corpus cannot be inferred for proprietary systems.

Taken together, our findings reveal several recurring weaknesses that can affect both the security and operational resilience of onboard satellite software.
One important result is that most findings occur in project-developed code, showing that most of the observed findings are located in code developed within the analyzed projects.
Looking at the aggregate distribution by weakness type, Memory Safety and Code Quality account for most of the findings across the dataset.
The concentration of incorrect argument types, integer overflows, uninitialized variables, and undefined behavior shows that many of the identified weaknesses affect the correctness and predictability of program execution.
These weaknesses can corrupt program state, alter control flow, or cause the software to behave unpredictably under unexpected inputs.
An attacker who can influence the inputs or conditions that trigger these weaknesses may induce behavior the software was not designed to handle, turning ordinary programming errors into opportunities to manipulate onboard functions, disrupt operations, or, in severe cases, damage the satellite.
Injection weaknesses also warrant particular attention.
These weaknesses can create a direct path from attacker-controlled inputs to sensitive onboard operations, allowing crafted inputs to reach command paths and cause unintended command execution or other unauthorized actions.
The SPARTA analysis further shows that the significance of these weaknesses depends on where they occur within the onboard architecture.
The concentration of Critical Subsystem Attack findings in ADCS and EPS, and of 3rd Party Software findings in payload software, indicates that security prioritization should consider subsystem function and potential mission impact in addition to generic severity and weakness category.

External dependency code presents an additional concern, accounting for 18.6\% of the identified findings.
This shows that a relevant portion of the observed findings comes from code developed and maintained outside the analyzed projects.
Addressing these weaknesses depends not only on identifying them, but also on tracking upstream fixes, evaluating their compatibility, and safely integrating updated versions into the onboard software stack.
In satellite systems, where software changes require extensive validation and post-launch update opportunities are limited, fixes may be difficult to adopt even after they become available.
This challenge is also reflected in our finding that no onboard software update capability was identified for most repositories in the corpus, potentially limiting the ability to remediate vulnerabilities after deployment.
External dependencies therefore require continuous inventory, version tracking, and upstream monitoring throughout development, integration, and mission operation.

The maintenance state of the repositories adds another layer of concern.
In our curation process, we found that 59 of the 126 repositories (46.8\%) had not been updated for more than two years at the time of analysis.
Stale repositories also showed a higher observed finding burden than maintained repositories, both overall and for High or Critical findings.
This lack of maintenance is particularly concerning in the context of space systems.
As new vulnerabilities are constantly being discovered, this lack of updates means known flaws can remain unaddressed in OSS repositories.
If other teams integrate this unmaintained OSS into their missions, these flaws can become embedded liabilities that persist for the multi-year or even decade-long lifespan of multiple missions.
It is therefore plausible that weaknesses similar to those identified in our study may also be present in software currently deployed on operational satellites, although the presence of such weaknesses in specific deployed configurations cannot be determined from our analysis.

The distribution of findings across repositories is also a key takeaway.
Security findings are widespread across the ecosystem but unevenly distributed, with 27.0\% of repositories containing no findings and 75.4\% containing 20 or fewer findings.
However, low counts do not necessarily imply low risk, particularly given that 71.8\% of the findings across the dataset were classified as medium severity or higher under the normalized severity scale.
Differences in finding counts may also reflect project-level characteristics such as repository size, project maturity, team size, development practices, and coding-standard compliance.
Assessing these factors consistently is challenging because it requires separating all the onboard-relevant software from other repository content and extracting comparable project-level information across heterogeneous projects.
Developing such an assessment could provide an additional perspective on the observed differences between repositories and is left for future work.
A structured discussion of construct, internal, external, and conclusion validity is provided in ~\ref{app:threats-validity}.
\section{Mitigation Strategies for Onboard OSS Security}

Mitigating the security risks associated with onboard OSS requires adapting established software-security practices to the operational constraints of satellite systems.
These practices should address the entire satellite software lifecycle, from planning and development through integration, pre-flight qualification, deployment, and in-orbit operation.
These strategies should account for restricted update opportunities, extensive verification requirements, operational resource constraints, and the strict reliability requirements of onboard systems.
Below, we outline several key mitigation strategies grounded in common OSS security practices and adapted for satellite systems.
Table~\ref{tab:mitigation_takeaways} summarizes the main takeaways and recommended actions.

\begin{table*}[t]
\centering
\small
\renewcommand{\arraystretch}{1.25}
\caption{Key takeaways and recommended actions for onboard OSS security.}
\label{tab:mitigation_takeaways}
\begin{tabularx}{\textwidth}{p{0.34\textwidth} X}
\toprule
\textbf{Key Takeaway} & \textbf{Recommended Action} \\
\midrule

Continuous vulnerability management is important for long-lived satellite software. &
Perform periodic SCA during development, pre-flight qualification, and mission operations, prioritize findings that affect deployed configurations, and pre-validate updates before deployment. \\

\addlinespace
Onboard software update support is not widespread. &
Treat post-deployment software update capability as an explicit design consideration and validate the update process before launch. \\

\addlinespace
External dependencies require continuous visibility. &
Maintain an up-to-date SBOM, compare newly disclosed vulnerabilities and dependency changes against the deployed flight-software baseline, and review dependency changes before approving new flight-software releases. \\

\addlinespace
Memory Safety and Code Quality require focused hardening. &
Use SAST, compiler diagnostics, code review, checked arithmetic, bounds validation, fuzzing, and negative testing before software is incorporated into an approved flight build. \\

\addlinespace
Input-facing and critical software require stronger containment. &
Apply strict input validation, least privilege, process isolation, recovery mechanisms, and build provenance controls to limit the impact of remaining weaknesses and support localized containment and recovery. \\

\bottomrule
\end{tabularx}
\end{table*}

\subsection{Proactive Vulnerability Monitoring and Patching}

A practical step toward improving the security of onboard OSS is to establish a vulnerability management process that continues throughout development and mission operations.
During development, teams can use SCA tools to identify known vulnerabilities in third-party dependencies and integrate their remediation into the software release process.
For satellites, this approach must account for limited uplink windows and safety-critical verification requirements~\cite{ncsc2022supplychain}, so that updates are bundled, pre-validated, and scheduled rather than applied ad hoc.

Because onboard software update support was not widespread across the analyzed corpus, post-deployment update capability should be treated as an explicit design consideration rather than an assumed operational feature.
Where supported by the mission architecture, the software update process should be defined and validated before launch so that identified vulnerabilities can be remediated during mission operation.

Operationally, we recommend running SCA periodically during development and pre-flight qualification, as well as at scheduled vulnerability review points during the mission.
Identified vulnerabilities should be prioritized according to whether they affect the deployed software configuration, their potential impact on mission-critical functions, and the availability of suitable remediation.
Aligning these activities with a structured secure development process, such as the NIST SSDF practices for identifying and addressing vulnerabilities~\cite{nistSSDF800218}, can reduce the likelihood that known issues persist throughout the satellite's lifetime.

\subsection{Software Supply Chain Security and SBOM Adoption}

Improving supply chain security for onboard OSS requires maintaining awareness of all third-party components included in the flight software.
Because many of these components are developed and maintained outside the mission team, operators must be able to identify their origin, version, dependency relationships, and maintenance status.
An SBOM should therefore be generated during development and maintained as part of the flight software baseline~\cite{ntiaSBOM2021}, providing a consistent record of the included dependencies and their versions.

The SBOM should be updated whenever a dependency is added, removed, or changed, and it should be reviewed before each software release is approved.
During mission operations, it provides the reference needed to determine whether newly disclosed vulnerabilities in upstream projects affect the exact software configuration deployed onboard.
By linking vulnerability disclosures to the deployed dependency set~\cite{nistSSDF800218}, the SBOM enables mission teams to assess whether action is required and to plan any necessary software changes through the mission's established verification and deployment process.
For long-lived missions, this assessment can be performed incrementally by comparing newly disclosed vulnerabilities and dependency changes against the maintained SBOM and deployed flight-software baseline.
This allows mission teams to focus their review on components whose security status has changed rather than repeatedly reassessing the complete software stack.
Such an approach is particularly useful when software modifications require extensive verification and re-qualification, as it allows the relevance of a newly disclosed issue to be established before initiating the more costly process of preparing and deploying an update.

\subsection{Memory-Safety and Code-Quality Hardening}

Because most findings occur in project-developed code, and Memory Safety and Code Quality account for most of the findings across the dataset, mitigation should also focus directly on secure implementation practices.
Mission teams should use strict compiler diagnostics, language-aware SAST, and code review to identify recurring weaknesses such as incorrect function arguments, integer overflows, uninitialized variables, unsafe memory operations, and undefined behavior before the software is incorporated into an approved flight build.
Because C and C++ analysis tools differ in the vulnerabilities they detect and may produce both false positives and false negatives~\cite{li2024evaluating}, findings affecting mission-critical code should be reviewed and resolved rather than relying on a single tool or automated result.

Beyond static analysis, teams should reduce opportunities for these weaknesses through defensive implementation and testing practices.
Code that handles buffers, pointers, numeric conversions, and externally influenced inputs should enforce explicit bounds, checked arithmetic, complete initialization, and clear ownership of memory and resources.
Sanitizer-enabled builds, coverage-guided fuzzing, and targeted negative testing~\cite{lee2026fuzzing} should also be applied during development and pre-flight qualification, particularly to command parsers, telemetry handlers, communication interfaces, configuration loaders, and other input-processing components.
Recent fuzzing approaches use code language models, reinforcement learning, and sequence-to-sequence models to improve vulnerability discovery and the exploration of complex program behaviors~\cite{yang2025fuzzcoder}.
Learning-guided input generation and adaptive exploration can help exercise execution paths that conventional mutation strategies may reach less effectively~\cite{yang2025dprfuzz}, while sequence-based models can further support the testing of structured inputs and protocol-oriented software~\cite{yang2024seq2seq}.
For onboard satellite software, these techniques could strengthen pre-flight testing of telemetry and telecommand parsers, protocol handlers, payload interfaces, and other stateful or structured input-processing components where malformed or unexpected inputs may expose fault paths that static analysis alone cannot evaluate.
Together, these practices can expose unsafe execution paths before deployment and reduce the likelihood that implementation weaknesses propagate into the qualified flight software.

\subsection{Standards-Based Input Handling, Isolation, and Build Integrity}

The prevalence of Injection, Insecure Design, and Mishandling of Exceptional Conditions findings indicates that onboard software should be designed to handle both externally influenced inputs and unexpected execution states safely.
Mission software should apply strict input validation, use parameterized or schema-driven interfaces, and avoid passing untrusted data to dynamic execution paths~\cite{owaspA052025}.
Focused CI rules should be used to identify injection-prone patterns, missing validation, unchecked error conditions, and unsafe resource handling~\cite{owaspA062025,owaspA102025,openssf2023} before software is incorporated into an approved flight build.
Together, these practices align with the controls recommended for Injection, Insecure Design, and Mishandling of Exceptional Conditions in the OWASP Top 10:2025, as well as with secure development guidance for OSS projects.

Mishandling of Exceptional Conditions and Insecure Design require additional attention to how onboard software is structured and how it behaves under both expected and failure conditions~\cite{owaspA062025,owaspA102025}.
Software requirements should define expected and failure states for critical operations, including responses to malformed commands, invalid values, unavailable resources, arithmetic errors, incomplete data, and resource exhaustion.
Negative testing, fuzzing, and fault-injection testing should be used to exercise these conditions and verify that the software rejects invalid operations, handles failures in a controlled manner, and transitions to an appropriate recovery or safe state~\cite{nasaSTD87398B}.
These practices provide a concrete means of identifying missing security controls and evaluating error-handling paths before the software is integrated into the flight system.
Satellite cybersecurity testbeds can support this process by enabling controlled evaluation of flight software under operational and adversarial conditions, while open-source and reproducible implementations enable repeatable experimentation and validation~\cite{idan2025aegissat,peled2025reproducible}.

At runtime, the potential impact of remaining weaknesses should be constrained through isolation and least privilege~\cite{nist80053r5}.
Where the onboard architecture permits, input-handling and externally exposed modules should be separated from mission-critical services, while resource permissions and inter-process communication should be restricted to the minimum necessary.
Resource limits, watchdog mechanisms, controlled process restarts, and predefined recovery procedures can further limit the propagation of abnormal behavior across the flight software~\cite{owaspA062025,owaspA102025}.
These controls limit the scope and impact of a triggered weakness and provide an additional layer of protection for weaknesses that remain after pre-flight analysis and testing.
This containment is particularly important for satellites because post-launch software updates may be constrained by communication opportunities and can require extensive verification and re-qualification before deployment.
Where supported by the onboard architecture, isolating input-facing components and enabling localized restart or recovery can contain a fault without requiring an immediate update of the complete flight-software stack.
These mechanisms do not replace vulnerability remediation, but can reduce operational exposure while an appropriate software update is prepared, validated, and deployed.

The build pipeline should apply similar principles of isolation and restricted trust.
Build steps should be isolated, the permissions granted to third-party dependencies and build scripts should be limited to the minimum necessary~\cite{nist204d}, and artifact provenance should be verified before software is promoted to approved flight branches.
Applying SLSA-aligned controls and software supply-chain practices in the CI pipeline~\cite{slsa2023} reduces the risk of unauthorized or unintended changes entering through third-party dependencies, build scripts, or compromised build processes, and supports the incorporation of approved and traceable artifacts into the flight software.
\section{Ethical Considerations}

The goal of this study is to identify recurring security weaknesses in OSS used in onboard satellite systems and help developers and mission teams address them.
Studying these weaknesses is important because issues in publicly available software may also affect software that is later integrated into satellite systems.
However, publishing the exact findings could make it easier to locate and misuse weaknesses in individual repositories.
For this reason, the study focuses on a high-level view of the findings and the broader trends across the repositories, without disclosing individual findings or linking them to specific repositories.
All analyses were performed offline on publicly available repositories, without interacting with operational satellites or deployed systems, and no identified weakness was exploited.
We recognize that withholding these details limits full finding-level reproducibility, but consider this necessary to reduce the risk of harm while still providing useful evidence about the security of onboard OSS.

\section{Conclusion}

This paper presented an empirical analysis of the security posture of OSS used in onboard satellite systems using a curated dataset of 126 public repositories and a multi-tool pipeline combining SBOM generation, SCA, SAST, IaC analysis, and secret scanning.
Most findings occurred in project-developed code, while external dependencies also accounted for a relevant portion of the findings.
Nearly half of the findings were attributed to OBC, CDH, and flight software.
The CWE analysis showed that Memory Safety and Code Quality accounted for most findings, while the OWASP-based analysis highlighted mishandling of exceptional conditions, insecure design, and injection as the main broader security-risk categories.
The SPARTA analysis further showed that the security context of these weaknesses differs across onboard subsystems, while onboard software update support was not widespread across the analyzed corpus.
Together, these results show that the observed security findings are concentrated primarily in project-developed code, while external dependencies also account for a relevant portion of the findings.

These results provide a basis for concrete actions by mission teams and the broader satellite software community.
For example, teams can maintain an SBOM as part of the flight software baseline and review it using SCA during development, pre-flight qualification, and scheduled mission reviews.
They can also use SAST, compiler diagnostics, checked arithmetic, bounds validation, and safer memory-handling practices to address recurring Memory Safety and Code Quality weaknesses.
Common injection patterns and secret leaks can be blocked in CI, while stronger input validation and error handling, isolation and least privilege for input-facing modules, and build provenance and dependency policies can help ensure that only reviewed artifacts reach approved flight branches.
Automated scans can produce false positives and do not establish runtime reachability or mission-specific exploitability, so findings should be validated using representative flight builds and the deployed software configuration.
By making security an integral part of the development and mission operations lifecycle, teams can strengthen the security of onboard OSS and make its risk posture easier to assess.
\section*{Funding}

This research did not receive any specific grant from funding agencies in the public, commercial, or not-for-profit sectors.

\section*{Data Availability}

The complete list of the 126 repositories analyzed in this study is provided in ~\ref{app:git-repos}.
The study was conducted between October 2025 and July 2026.
The finding-level dataset is not publicly available because it contains security-sensitive information that could be used to identify weaknesses in individual repositories.

\section*{Declaration of Generative AI Use}

During the preparation of this manuscript, the authors used OpenAI's ChatGPT and Anthropic's Claude to assist with language review, grammar correction, and editorial refinement.
These tools were not used to generate or analyze research data, produce experimental results, or create figures.
After using these tools, the authors reviewed and edited the content as needed and take full responsibility for the content of the manuscript.

\bibliographystyle{elsarticle-num}
\bibliography{references}

\appendix
\appendix

\section{Git Repositories Used}
\label{app:git-repos}

This appendix lists the public Git repositories analyzed in this study.
The repositories are sorted alphabetically by repository name.
The full list is provided in Tables~\ref{tab:oss_repos_1}--\ref{tab:oss_repos_4}.

\begin{table*}[p]
\centering
\scriptsize
\setlength{\tabcolsep}{5pt}
\renewcommand{\arraystretch}{1.05}
\caption{Public repositories analyzed in this study (Part 1 of 4).}
\label{tab:oss_repos_1}
\begin{tabularx}{\textwidth}{p{0.27\textwidth} X}
\toprule
\textbf{Repository} & \textbf{Link} \\
\midrule
1KCubeSat\_Software & \url{https://github.com/rgw3d/1KCubeSat_Software.git} \\
ACRUX-2 Flight Software & \url{https://github.com/MelbourneSpaceProgram/acrux-flight-software} \\
ACS & \url{https://github.com/Alpha-CubeSat/ACS.git} \\
adcs-on-board-software & \url{https://github.com/AcubeSAT/adcs-on-board-software.git} \\
adcs\_firmware & \url{https://github.com/cmattatall/adcs_firmware.git} \\
AmbaSat-1 & \url{https://github.com/ambasat/AmbaSat-1.git} \\
artemis-cubesat & \url{https://github.com/hsfl/artemis-cubesat} \\
Big-Red-Sat-1 & \url{https://github.com/Big-Red-Sat/Big-Red-Sat-1} \\
BIRDS3-OBC & \url{https://github.com/BIRDSOpenSource/BIRDS3-OBC} \\
BIRDS4-OBC & \url{https://github.com/BIRDSOpenSource/BIRDS4-OBC} \\
boot\_flight\_software & \url{https://github.com/MelbourneSpaceProgram/boot_flight_software} \\
bp & \url{https://github.com/nasa/bp} \\
c2a-aobc & \url{https://github.com/ut-issl/c2a-aobc} \\
c2a-core & \url{https://github.com/ut-issl/c2a-core.git} \\
c2a-user-for-raspi & \url{https://github.com/arkedge/c2a-user-for-raspi.git} \\
canopus & \url{https://github.com/satellogic/canopus.git} \\
CF & \url{https://github.com/nasa/CF} \\
cFE & \url{https://github.com/nasa/cFE.git} \\
cFS & \url{https://github.com/nasa/cFS.git} \\
CHESS Flight Software & \url{https://github.com/CHESS-mission/05_FS} \\
coconut-fsw & \url{https://github.com/ASU-SDSL/coconut-fsw} \\
coreflightexec & \url{https://github.com/junitas/coreflightexec.git} \\
CryptoLib & \url{https://github.com/nasa/CryptoLib} \\
CS & \url{https://github.com/nasa/CS.git} \\
Cubedate & \url{https://github.com/thingsat/Cubedate.git} \\
CubeSat & \url{https://github.com/cubesat-project/CubeSat.git} \\
CubeSat & \url{https://github.com/saivan/CubeSat.git} \\
CubeSat-ADCS-old & \url{https://github.com/DalhousieSpaceSystemsLab/CubeSat-ADCS-old.git} \\
CubeSat-Camera-Interface & \url{https://github.com/DalhousieSpaceSystemsLab/CubeSat-Camera-Interface.git} \\
CubeSat-Core & \url{https://github.com/DalhousieSpaceSystemsLab/CubeSat-Core.git} \\
CubeSat-Reaction-Wheel-control & \url{https://github.com/yiqiangjizhang/CubeSat-Reaction-Wheel-control.git} \\
CubeSat-v3 & \url{https://github.com/windymsth/CubeSat-v3.git} \\
\bottomrule
\end{tabularx}
\end{table*}

\begin{table*}[p]
\centering
\scriptsize
\setlength{\tabcolsep}{5pt}
\renewcommand{\arraystretch}{1.05}
\caption{Public repositories analyzed in this study (Part 2 of 4).}
\label{tab:oss_repos_2}
\begin{tabularx}{\textwidth}{p{0.27\textwidth} X}
\toprule
\textbf{Repository} & \textbf{Link} \\
\midrule
CubeSatADCS & \url{https://github.com/rachit31/CubeSatADCS.git} \\
CubeSatADCS & \url{https://github.com/RemyChatel/CubeSatADCS.git} \\
CubeWorks & \url{https://github.com/SmallSatGasTeam/CubeWorks} \\
CySat 1 & \url{https://github.com/M2I-CySat/CySat-1-Main.git} \\
DARS & \url{https://github.com/AlaskaResearchCubeSat/DARS.git} \\
DS & \url{https://github.com/nasa/DS.git} \\
DubSat1 & \url{https://github.com/UWCubeSat/DubSat1.git} \\
ecss-services & \url{https://github.com/AcubeSAT/ecss-services.git} \\
eps2 & \url{https://github.com/spacelab-ufsc/eps2} \\
Femto-satellite & \url{https://github.com/spel-uchile/Femto-satellite.git} \\
finch-firmware & \url{https://github.com/utat-ss/finch-firmware} \\
firmware & \url{https://github.com/pycubed/firmware.git} \\
floripasat & \url{https://github.com/floripasat/floripasat.git} \\
FM & \url{https://github.com/nasa/FM.git} \\
FOD & \url{https://github.com/spel-uchile/FOD.git} \\
FOSSASAT-1 & \url{https://github.com/FOSSASystems/FOSSASAT-1} \\
FOSSASAT-1B & \url{https://github.com/FOSSASystems/FOSSASAT-1B} \\
FOSSASAT-2 & \url{https://github.com/FOSSASystems/FOSSASAT-2} \\
fprime-artemis-cubesat-c3m & \url{https://github.com/hsfl/fprime-artemis-cubesat-c3m} \\
fsp & \url{https://github.com/floripasat/fsp.git} \\
generic\_eps & \url{https://github.com/nasa-itc/generic_eps} \\
generic\_radio & \url{https://github.com/nasa-itc/generic_radio} \\
generic\_star\_tracker & \url{https://github.com/nasa-itc/generic_star_tracker} \\
generic\_torquer & \url{https://github.com/nasa-itc/generic_torquer} \\
GlobusSatProject2 & \url{https://github.com/adiwwww/GlobusSatProject2} \\
HDTN & \url{https://github.com/nasa/HDTN.git} \\
Hermes & \url{https://github.com/ThomasMontano/Hermes.git} \\
hestia & \url{https://github.com/mawson-rovers/hestia} \\
HK & \url{https://github.com/nasa/HK.git} \\
HS & \url{https://github.com/nasa/HS.git} \\
IntelliSat & \url{https://github.com/Space-and-Satellite-Systems-UC-Davis/IntelliSat.git} \\
KS-1Q & \url{https://github.com/opensatellite/KS-1Q.git} \\
\bottomrule
\end{tabularx}
\end{table*}

\begin{table*}[p]
\centering
\scriptsize
\setlength{\tabcolsep}{5pt}
\renewcommand{\arraystretch}{1.05}
\caption{Public repositories analyzed in this study (Part 3 of 4).}
\label{tab:oss_repos_3}
\begin{tabularx}{\textwidth}{p{0.27\textwidth} X}
\toprule
\textbf{Repository} & \textbf{Link} \\
\midrule
LC & \url{https://github.com/nasa/LC.git} \\
libcsp & \url{https://github.com/libcsp/libcsp.git} \\
libraries & \url{https://github.com/cubesatplatform/libraries.git} \\
lost & \url{https://github.com/UWCubeSat/lost.git} \\
low-gain-radio & \url{https://github.com/oresat/low-gain-radio.git} \\
MD & \url{https://github.com/nasa/MD} \\
MM & \url{https://github.com/nasa/MM.git} \\
MSP Flight Software & \url{https://github.com/MelbourneSpaceProgram/msp_flight_software_public} \\
nanosat-mo-framework & \url{https://github.com/esa/nanosat-mo-framework.git} \\
NASA F Prime & \url{https://github.com/nasa/fprime} \\
NEAScout-Science & \url{https://github.com/JPLMLIA/NEAScout-Science} \\
novatel\_oem615 & \url{https://github.com/nasa-itc/novatel_oem615} \\
NYUSat & \url{https://github.com/EasonNYC/NYUSat} \\
obc-firmware & \url{https://github.com/SFUSatClub/obc-firmware.git} \\
obc-software & \url{https://github.com/AcubeSAT/obc-software.git} \\
obdh & \url{https://github.com/floripasat/obdh.git} \\
obdh2 & \url{https://github.com/spacelab-ufsc/obdh2.git} \\
oop-chipsat-code & \url{https://github.com/Alpha-CubeSat/oop-chipsat-code.git} \\
oop-flight-code & \url{https://github.com/Alpha-CubeSat/oop-flight-code.git} \\
open-set & \url{https://github.com/jaxa/open-set} \\
OpenSatKit & \url{https://github.com/OpenSatKit/OpenSatKit.git} \\
opssat-saasy-ml & \url{https://github.com/visionspacetec/opssat-saasy-ml.git} \\
opssat-smartcam & \url{https://github.com/georgeslabreche/opssat-smartcam.git} \\
orbital-platform & \url{https://github.com/uwu64/orbital-platform.git} \\
oresat-adcs-software & \url{https://github.com/oresat/oresat-adcs-software} \\
oresat-c3-software & \url{https://github.com/oresat/oresat-c3-software} \\
oresat-firmware & \url{https://github.com/oresat/oresat-firmware} \\
oresat-gps-software & \url{https://github.com/oresat/oresat-gps-software} \\
oresat-linux & \url{https://github.com/oresat/oresat-linux} \\
osal & \url{https://github.com/nasa/osal.git} \\
pfl & \url{https://github.com/smorad/pfl.git} \\
pfs & \url{https://github.com/TJREVERB/pfs.git} \\
\bottomrule
\end{tabularx}
\end{table*}

\begin{table*}[p]
\centering
\scriptsize
\setlength{\tabcolsep}{5pt}
\renewcommand{\arraystretch}{1.05}
\caption{Public repositories analyzed in this study (Part 4 of 4).}
\label{tab:oss_repos_4}
\begin{tabularx}{\textwidth}{p{0.27\textwidth} X}
\toprule
\textbf{Repository} & \textbf{Link} \\
\midrule
PHASMA OBC Software & \url{https://gitlab.com/librespacefoundation/phasma/phasma-obc-software} \\
pi-sat & \url{https://github.com/OpenSatKit/pi-sat.git} \\
PSP & \url{https://github.com/nasa/PSP.git} \\
qubesat & \url{https://github.com/space-technologies-at-california/qubesat.git} \\
Quetzal-1 software & \url{https://github.com/Quetzal-1-CubeSat-Team/quetzal1-flight-software.git} \\
samwise-adcs-flight & \url{https://github.com/stanford-ssi/samwise-adcs-flight} \\
samwise-flight-software & \url{https://github.com/stanford-ssi/samwise-flight-software} \\
sat-rs & \url{https://github.com/us-irs/sat-rs} \\
satllazero & \url{https://github.com/kcglab/satllazero} \\
SatOS-Payload-SDK & \url{https://github.com/antaris-inc/SatOS-Payload-SDK.git} \\
SBN & \url{https://github.com/nasa/SBN} \\
SC & \url{https://github.com/nasa/SC.git} \\
SCALES F Prime Reference & \url{https://github.com/BroncoSpace-Lab/fprime-scales-ref} \\
SCH & \url{https://github.com/nasa/SCH.git} \\
scsat1-fsw & \url{https://github.com/spacecubics/scsat1-fsw} \\
scsat1-rpi & \url{https://github.com/spacecubics/scsat1-rpi} \\
seaking1 & \url{https://github.com/cubesatplatform/seaking1.git} \\
shflight & \url{https://github.com/SPACE-HAUC/shflight.git} \\
soci\_cdh\_rtos & \url{https://github.com/AA-CubeSat-Team/soci_cdh_rtos} \\
software & \url{https://github.com/pycubed/software.git} \\
Star\_Tracker & \url{https://github.com/spel-uchile/Star_Tracker.git} \\
SUCHAI-Flight-Software & \url{https://gitlab.com/spel-uchile/suchai-flight-software} \\
ttc & \url{https://github.com/floripasat/ttc.git} \\
ttc2 & \url{https://github.com/spacelab-ufsc/ttc2} \\
upsat-adcs-software & \url{https://gitlab.com/librespacefoundation/upsat/upsat-adcs-software} \\
upsat-comms-software & \url{https://gitlab.com/librespacefoundation/upsat/upsat-comms-software} \\
upsat-ecss-services & \url{https://gitlab.com/librespacefoundation/upsat/upsat-ecss-services} \\
upsat-eps-software & \url{https://gitlab.com/librespacefoundation/upsat/upsat-eps-software} \\
upsat-iac-software & \url{https://gitlab.com/librespacefoundation/upsat/upsat-iac-software} \\
upsat-obc-software & \url{https://gitlab.com/librespacefoundation/upsat/upsat-obc-software} \\
\bottomrule
\end{tabularx}
\end{table*}

\clearpage

\section{Security Analysis Tool Versions and Parameters}
\label{app:tool-config}

Table~\ref{tab:tool_versions} reports the versions used for each security analysis tool.

\begin{table}[t]
\centering
\small
\caption{Latest versions of the security analysis tools used in the study.}
\label{tab:tool_versions}
\begin{tabular}{ll}
\toprule
\textbf{Tool} & \textbf{Version} \\
\midrule
Syft                   & 1.33.0 \\
Grype                  & 0.100.0 \\
Trivy                  & 0.52.2 \\
Semgrep                & 1.139.0 \\
Gitleaks               & 8.28.0 \\
OSV-Scanner            & 2.2.3 \\
OWASP Dependency-Check & 12.1.6 \\
Bandit                 & 1.9.4 \\
Checkov                & 3.2.513 \\
Bearer                 & 1.51.0 \\
Cppcheck               & 2.13.0 \\
CodeQL                 & 2.25.5 \\
\bottomrule
\end{tabular}
\end{table}

The tools were executed using their standard repository or recursive scan modes.
Semgrep used the \texttt{auto} ruleset, Trivy enabled vulnerability, secret, and misconfiguration scanning, Cppcheck enabled all checks, and CodeQL used its default query selection.
Bandit and Cppcheck were applied when the corresponding Python or C/C++ source files were present, while Checkov was applied when supported configuration files were available.

\section{Dataset Filtering Breakdown}
\label{app:filtering-breakdown}

Table~\ref{tab:filtering_breakdown} summarizes the number and proportion of findings removed from the 28,274 candidate findings for each filtering reason, resulting in the final dataset of 2,827 findings.

\begin{table*}[t]
\centering
\small
\renewcommand{\arraystretch}{1.15}
\caption{Net breakdown of findings eliminated during dataset construction. Percentages are relative to the 28,274 candidate findings.}
\label{tab:filtering_breakdown}
\begin{tabularx}{\textwidth}{Xrr}
\toprule
\textbf{Elimination reason}
& \textbf{Findings removed}
& \textbf{\%} \\
\midrule

Parser and path-based exclusions
& 9,380 & 33.18 \\

Third-party or platform code outside the onboard configuration
& 5,661 & 20.02 \\

Ground-side and mission-operations software
& 2,671 & 9.45 \\

Non-security, noisy, or redundant scanner findings
& 2,627 & 9.29 \\

Build, CI, packaging, or generated artifacts
& 1,458 & 5.16 \\

Mixed non-onboard paths
& 1,444 & 5.11 \\

Simulator- or emulation-only code
& 724 & 2.56 \\

Tests, examples, demos, or benchmarks
& 570 & 2.02 \\

Duplicate detections
& 539 & 1.91 \\

Other code outside the onboard scope
& 268 & 0.95 \\

Final manual scope exclusions
& 54 & 0.19 \\

Documentation and generated reports
& 51 & 0.18 \\

\midrule
\textbf{Total eliminated}
& \textbf{25,447}
& \textbf{90\%} \\

\textbf{Final retained dataset}
& \textbf{2,827}
& \textbf{10\%} \\

\bottomrule
\end{tabularx}
\end{table*}

\section{Threats to Validity}
\label{app:threats-validity}

This section discusses threats to construct, internal, external, and conclusion validity.
Construct validity is primarily related to the use of findings produced by automated security-analysis tools to characterize recurring security patterns across onboard-relevant OSS.
To address this, we reviewed a random sample of 360 retained findings to estimate the false-positive rate and analyzed the dataset separately by finding class and reporting tool.
Internal validity may be affected by repository-scope classification, filtering, deduplication, and normalization decisions.
All retained findings were reviewed for onboard relevance, a separate sample of excluded findings was examined for incorrect removal, and the filtering and classification process was maintained through an auditable analysis pipeline.

External validity is determined by the scope of the analyzed corpus, which consists of publicly available OSS used in or developed for onboard satellite systems.
The observed patterns therefore characterize this corpus, which includes both software used in space and software intended for future onboard integration, while proprietary flight software is outside the scope of the study.
Conclusion validity may be affected by differences in tool applicability, language support, and reporting behavior across finding types.
We address these factors through finding-class and tool-specific analyses, sensitivity analysis of the aggregate results, and separate severity analysis across tools and finding classes.
The false-positive estimate is additionally reported with a 95\% confidence interval to quantify the uncertainty associated with the sampled validation.

\clearpage
\onecolumn

\noindent
\begin{minipage}[t]{0.48\textwidth}
\vspace{0pt}

\section{Severity Normalization Mapping}
\label{app:severity-mapping}

Because the analysis tools use heterogeneous severity representations, severity normalization followed an evidence-precedence procedure based on the strongest applicable severity information available for each finding.
When available, a CVSS-based score associated with a CVE or vulnerability advisory was used as the highest-priority severity evidence.
When such a score was unavailable, tool-defined numeric security scores and then categorical severity information were used.
Table~\ref{tab:severity-mapping} summarizes the three priority levels used to map this information to the common four-level severity scale.
Following the initial normalization, documented rule- and location-specific corrections were applied where the recorded finding context required an adjustment.

For the sensitivity analysis, each of the 2,827 retained findings was classified into one of five finding classes based on its evidentiary basis: 1,330 security-focused source-code alerts, 1,157 correctness and quality warnings, 263 vulnerable-dependency or advisory matches, 42 secret candidates, and 35 configuration or IaC findings.
Cppcheck was applicable to 113 of the 126 repositories, reflecting the prevalence of C/C++ source code across the analyzed corpus.
Severity varied across these classes and tools: medium-or-higher findings accounted for 86.0\% of security-focused source-code alerts and 49.4\% of correctness and quality warnings, while the corresponding proportion was 60.2\% for Cppcheck findings and 84.6\% for findings reported by the remaining tools.

\section{Lower-Frequency SPARTA Threats to Space Systems}
\label{app:sparta_threats}

The remaining nine SPARTA Threats to Space Systems categories each contain fewer than 20 findings and together account for 83 of the 2,827 retained findings (2.9\%).
Figure~\ref{fig:sparta_threat_tail} presents their distribution across the onboard software subsystems.
The bars show the total number of findings associated with each threat category and their distribution across the corresponding subsystems.

Exploit Lack of Bus Segregation is the largest of these categories with 16 findings, 14 of which occur in payload and mission software.
Software Defined Radio and Inadequate Security Planning/Design each account for 11 findings and are also concentrated in payload and mission software.
On-Orbit Software Update accounts for 10 findings, nine of which occur in OBC, CDH, and flight software, while Fault Management Exploitation and Malicious Use Of Hardware Commands contain seven and six findings, respectively, all within this subsystem.
Unauthorized Access contains nine findings, all in shared runtime software, while Compromise/Corrupt Running State and Timing Disruption are distributed across multiple onboard subsystems.

\end{minipage}
\hfill
\begin{minipage}[t]{0.48\textwidth}
\vspace{0pt}

\captionsetup{hypcap=false}

\begingroup
\renewcommand{\thetable}{E.\arabic{table}}

\captionof{table}{Evidence-precedence rules used to normalize severity information to the common four-level scale.}
\label{tab:severity-mapping}

\endgroup

\centering
\footnotesize
\setlength{\tabcolsep}{3pt}
\renewcommand{\arraystretch}{1.12}

\begin{tabular}{p{0.12\linewidth} p{0.30\linewidth} p{0.46\linewidth}}
\hline
\textbf{Priority} & \textbf{Severity evidence} & \textbf{Normalization rule} \\
\hline

1
& CVSS-based score associated with a CVE or vulnerability advisory
& Critical: $\geq9.0$; High: $\geq7.0$; Medium: $\geq4.0$; Low: $<4.0$ \\

2
& Tool-defined numeric security score
& Mapped to the corresponding Critical, High, Medium, or Low level according to the severity definition associated with the reported numeric score \\

3
& Categorical severity information
& Native categorical severity values were mapped to the corresponding Critical, High, Medium, or Low level.
Categorical \texttt{critical} was mapped to Critical, \texttt{high} and severity-level \texttt{error} to High, \texttt{medium} and \texttt{warning} to Medium, and \texttt{low}, \texttt{info}, \texttt{recommendation}, and \texttt{note} to Low.
For code-analysis categories consisting of \texttt{error}, \texttt{warning}, \texttt{style}, \texttt{performance}, and \texttt{portability}, \texttt{error} and \texttt{warning} were mapped to Medium, while \texttt{style}, \texttt{performance}, and \texttt{portability} were mapped to Low. \\

\hline
\end{tabular}

\vspace{1.2em}

\includegraphics[
    width=\linewidth
]{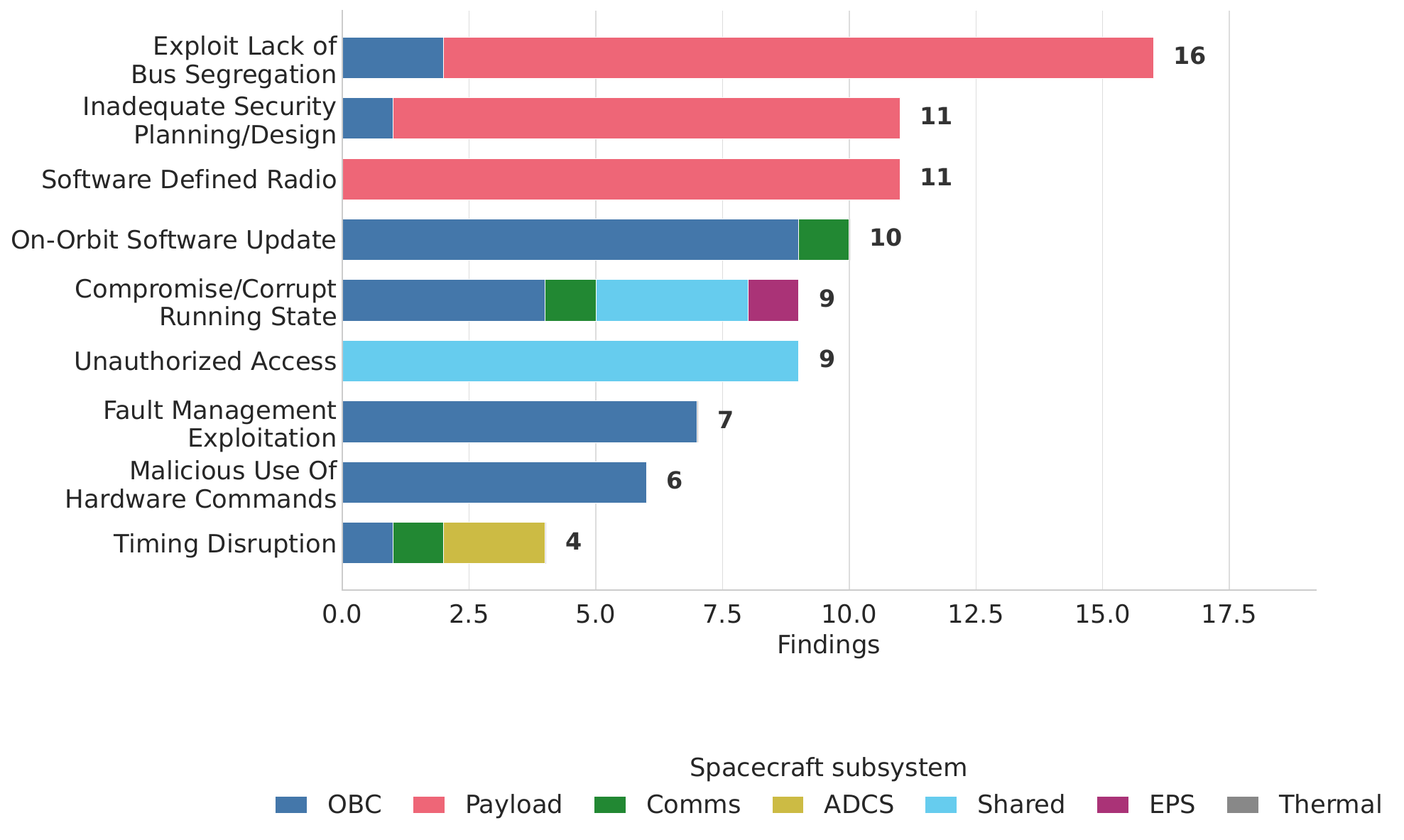}

\captionof{figure}{Distribution of the lower-frequency SPARTA Threats to Space Systems categories across onboard software subsystems. Each bar shows the number of findings associated with the corresponding threat category and their subsystem distribution.}
\label{fig:sparta_threat_tail}

\end{minipage}

\end{document}